 \documentclass[final,5p,times,twocolumn,authoryear]{elsarticle}

\usepackage{amssymb}
\usepackage{amsmath}
\usepackage{lipsum}
\usepackage[colorlinks = true, linkcolor = blue, urlcolor  = blue, citecolor = blue]{hyperref}
\usepackage{graphicx}
\usepackage{subcaption}
\usepackage{float}
\usepackage{todonotes}

\newcommand{\aste}{{\tt ASTErIsM}}
\newcommand{\dru}{{\tt DRUID}}
\newcommand{\se}{{\tt SourceExtractor++}}
\newcommand{\scarlet}{{\tt Scarlet}}
\newcommand{\unet}{{\tt U-Net}}
\newcommand{\CNN}{{\tt Mask R-CNN}}
\newcommand{\ADP}{{\tt ADP}}
\newcommand{\DADP}{{\tt DADP}}
\newcommand{\DBSCAN}{{\tt DBSCAN}}
\newcommand{\DENCLUE}{{\tt DENCLUE}}

\journal{Astronomy $\&$ Computing}

\begin{document}

\begin{frontmatter}



\title{Fast and accurate astronomical source deblending with Density-Peak Clustering}


\author[OATS]{M. D. Lepinzan}

\author[U_TS,OATS]{F. Tomba}

\author[OATS]{E. Romelli}

\author[U_TS]{A. Rodriguez}

\author[OATS]{L. Tornatore}

\affiliation[OATS]{organization={INAF-Osservatorio Astronomico di Trieste},
            addressline={via Tiepolo 11}, 
            city={Trieste},
            postcode={34143}, 
            state={TS},
            country={Italy}}
\affiliation[U_TS]{organization={Dipartimento di Matematica, Sezione di Matematica e Geoscienze, Università di Trieste},
            addressline={via Weiss 2}, 
            city={Trieste},
            postcode={34128}, 
            state={TS},
            country={Italy}}

\begin{abstract}
Source deblending represents a fundamental challenge for current and forthcoming astronomical surveys, where the increasing source density and image depth lead to a growing number of overlapping detections. Accurate deblending is essential for reliable measurements of source morphology and photometric properties, as well as for cosmological analyses based on the spatial distribution and shapes of galaxies. 

In this work, we present a redesign of the Advanced Density Peak (\ADP) clustering algorithm, tailored to the identification and separation of blended astronomical sources within detection regions. We develop a modular validation framework combining realistic image simulations, automatically generated ground-truth segmentation, and label-invariant evaluation metrics. The performance of \ADP\ is assessed against \aste, an established density-based astronomical deblender, through controlled pairwise simulations, synthetic multi-source images, and observations from the \textit{Euclid} Q1 public data release.

In pairwise simulations, \ADP\ and \aste\ show comparable deblending performance across a broad range of source separations and flux ratios, with photometric differences typically below 1\%, and deviations reaching at most $5-7\%$ in the most challenging configurations. No systematic bias is observed, as both algorithms alternately yield larger or smaller flux estimates depending on the specific blend geometry. In multi-source simulations, \ADP\ recovers approximately 8\% more ground-truth sources than \aste, while the positions of sources identified by both algorithms agree at the sub-pixel level. On \textit{Euclid} Q1 public data release, the two methods show strong agreement in segmentation area, ellipticity, position angle, and photometry, with the largest differences occurring for the smallest and faintest sources.

\ADP\ also provides a substantial computational advantage. End-to-end benchmarks on $19200\times19200$ pixels \textit{Euclid} Q1 public data release require approximately 16--200~s, corresponding to speedups of $12\times$--$156\times$ relative to \aste, with median and mean improvements of $36\times$ and $54\times$, respectively. These results show that \ADP\ provides scientifically competitive deblending at substantially lower computational cost, making it a promising approach for large-scale astronomical imaging surveys.

\end{abstract}




\end{frontmatter}




\section{Introduction}
\label{introduction}

The accurate identification and separation of astronomical sources from imaging data, commonly referred to as source deblending, is a fundamental task in modern observational astronomy. With the advent of wide-field surveys and high-resolution instruments such as \textit{Euclid}~\citep{Euclid:2024yrr}, Large Synoptic Survey Telescope~\citep[LSST,][]{LSST:2008ijt} and James Webb Space Telescope~\citep[JWST,][]{McElwain_2023}, the density and level of detail of objects captured in a single image has dramatically increased, often resulting in overlapping or closely spaced sources. For instance, the public \textit{Euclid} Quick Data Release~\citep[EQ1,][]{Euclid:2025rvk} already contains over 30 million sources across just $\sim63$ $\mathrm{deg^2}$ of sky, and by the end of the mission the survey is expected to cover roughly 14000 $\mathrm{deg^2}$. 

This blending of sources poses a significant challenge for downstream tasks such as photometry, morphological classification, and object cataloging~\citep{Bosch2018, Euclid:2025wfi}. Accurate deblending is therefore essential for the reliable recovery of source photometric and morphological properties, with implications extending to cosmological analyses based on the spatial distribution and shapes of galaxies. Missed, merged, or incorrectly deblended sources can alter the observed galaxy population and introduce systematic effects in galaxy-clustering and weak-lensing measurements, potentially propagating into cosmological inference~\citep{Melchior:2021bdg, Levine:2024svv}.

To address this, various deblending algorithms have been developed, ranging from threshold-based techniques~\citep[\se;][]{2020ASPC..527..461B, Kummel2022}, to model-fitting methods~\citep[\scarlet;][]{Melchior2018}, to density-based approaches~\citep[\aste;][]{tramacere2016ASTErIs}, and to topological methods based on persistent homology ~\citep[\dru;][]{2025RASTI...4....6S}. Each approach trades accuracy, robustness, and computational cost differently: thresholding is fast but struggles in crowded fields; model fitting can be precise but is computationally heavy and sensitive to initial conditions; density and topological approaches are more agnostic to morphology but often require careful scale choices. 

Outside astronomy, neural network-based methods nowadays represent the state of the art. Models such \unet~\citep{ronneberger2015unet} and its variants have been proven to handle successfully segmentation tasks with complex decision boundaries. However, their use has so far been constrained to relatively small image patches ($\approx 128 \times 128$ pixels). More complex models like \CNN~\citep{burke2019mask-r-cnn} can overcome patch-size limitations, but at the expense of substantial computational cost in both training and inference, and without guaranteeing the transferability of model accuracy on data coming from different sources. 

Despite these advances, two key challenges remain: the absence of systematic benchmarks that stress-test deblending methods under controlled simulations, for example by varying flux-ratio and source separation, and the computational cost associated with processing the tens of millions of sources expected from surveys such as \textit{Euclid}. In this context, reducing deblending runtime from hours to minutes can substantially decrease the computational resources required and facilitate the subsequent stages of the analysis pipeline.

In this work, we present a novel adaptation of the Advanced Density Peak~\citep[\ADP,][]{derrico2021adp} clustering algorithm for the segmentation of blended sources in astronomical images. Our method operates directly on pixel-level flux and requires no assumptions about source morphology. To ensure survey-scale efficiency and avoid deblending becoming a bottleneck in large data-processing pipelines, we also present as a future development a distributed-memory implementation of the algorithm called Distributed Advanced Density Peak (\DADP), which will enable scalability to the data volumes of current and upcoming wide-field surveys.

To rigorously test \ADP\, we develop a modular validation framework combining controlled simulations of point-like and extended sources, physically motivated ground-truth (GT) segmentation maps~\citep{haigh2021optimising}, and label-invariant evaluation metrics. The validation is organized at two complementary levels: controlled pairwise simulations are used to characterize the deblending behavior as a function of source separation and flux ratio, while synthetic multi-source images provide a more realistic environment for assessing source recovery and segmentation in complex configurations. This approach enables both the accuracy of the recovered segmentation and the occurrence of over and under-deblending to be quantified against a known reference. Although developed to validate \ADP, the framework is independent of the underlying deblending method and can therefore be readily applied to benchmark alternative algorithms.

The observational validation is framed within the \textit{Euclid} context, whose high source density and stringent requirements on source detection and characterization make deblending particularly challenging. The validation is therefore extended to real observations from the EQ1 public data release, where \ADP\ is compared with \aste\ through the morphological and photometric properties of the recovered sources. \aste\ provides a particularly relevant benchmark, as it is an established density-based deblending algorithm specifically developed for astronomical images and was used to process crowded fields included in the EQ1 public release. Its adoption therefore provides a well-established and observationally relevant reference against which the performance of \ADP\ can be assessed on the same data. Finally, we assess the computational performance of \ADP\ through strong and weak-scaling tests of its OpenMP implementation and an end-to-end runtime comparison with \aste\ on a sample of EQ1 public data release.

This works is organized as follows. Section~\ref{ADP_description} describes the \ADP\ algorithm and its adaptation to astronomical source deblending, together with its distributed-memory implementation, \DADP, designed to extend the patch-based approach beyond a single compute node. Section~\ref{Validation_pipeline} introduces the validation and performance methodology, including controlled simulations, GT construction, EQ1 public data release comparison, and computational benchmarks. The corresponding results are presented in Section~\ref{Results}, where \ADP\ is compared with \aste\ on simulated and EQ1 data and its computational performance is assessed. Finally, Section~\ref{conclusion} summarizes the main findings and discusses future developments.

\section{From density-peak clustering to astronomical deblending}
\label{ADP_description}
\begin{figure*}[t]
    \centering
    \includegraphics[width=1.0\textwidth]{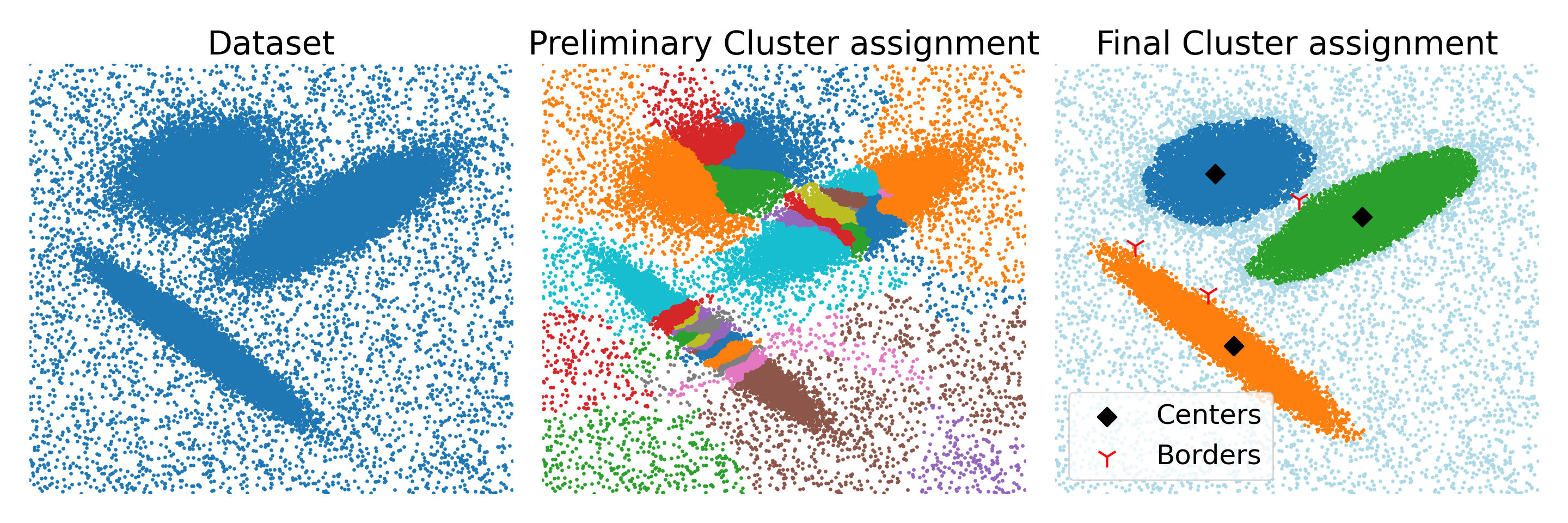}
   
    \caption{Example of \ADP\ applied to a synthetic dataset. The left panel shows a mixture of four Gaussians distributions with additional uniform noise. The middle panel depicts the initial clustering before the density peak merging validation step. The right panel displays the final clusters obtained after the merging procedure.}
    \label{fig:ADP_procedure}
\end{figure*}
The deblending  problem can be formally cast as an image segmentation task, where the goal is to partition a detected region into distinct sources by identifying local maxima of the luminosity field and assigning the surrounding pixels to the corresponding components. This formulation naturally connects with density-based clustering, in which clusters are identified as regions associated with local density peaks in a point cloud.

Several astronomical applications have already leveraged this connection. In particular, \aste\ combines two well-established density-based clustering techniques, \DBSCAN~\citep{ester1996proc} and \DENCLUE~\citep{Hinneburg1998AnEA,hinneburg2007denclue}, for source detection and deblending in astronomical images, and it is currently adopted within the \textit{Euclid} data processing pipeline. While both algorithms provide a well-established framework for density-based clustering, they present some limitations. \DBSCAN~is highly sensitive to the choice of its parameters, and can struggle when clusters exhibit substantially different densities, whereas the iterative nature of \DENCLUE, can result in a significant computational cost. These limitations can propagate to their application in astronomical deblending, motivating the exploration of alternative density-based approaches that retain the advantages of this framework while improving flexibility and computational performance.

In the realm of density-based clustering methods, \ADP\ stands out as one of the most recent developments, combining the density-based framework with hierarchical structures detection. These features make it a compelling candidate for addressing the deblending problem. 
In its original formulation, the \ADP\ algorithm requires as input a dataset $\mathbf{X}= \{\mathbf{x_1}, \dots, \mathbf{x_n}\}$ with $\mathbf{x_i} \in \mathbf{R}^n$. For each data point, an estimate of the log-density value $\log(\rho_i)$ and its associated error $\varepsilon_i$ are required, together with the list of its \textit{k-nearest} neighbors. The clustering procedure consists of four main steps.

\paragraph{1) Estimate of the local density and the associated error} 
The preliminary step consists in organizing the data so that the \textit{k-nearest} neighbors problem can be easily solved. In low-dimensional data setups, this is achieved using a spatial indexing structure like a \textit{kd-tree}~\citep{bentley1975kdtree}, a \textit{ball-tree}~\citep{omohundro1989balltree} or a \textit{Bounding Volume Hierarchy}~\citep[BVH,][]{clark1976bvh,rubin1980bvh}. For high-dimensional data, the only solution is a brute force approach. Once the \textit{k-nearest} neighbors for each point have been found, they are used to estimate the density value at each point. In this case \ADP~requires an estimator that returns both the estimate of the density value and the associated error. These can be, for example, the \textit{Point Adaptive k}~\citep[PAk,][]{rodriguez2018computing} estimator and the \textit{Binless Multidimensional Thermodynamic Integration}~\citep[BMTI,][]{carli2024density} estimator.

\paragraph{2) Cluster center identification} This step identifies cluster centers by searching for local maxima of the density field. Once the centers are determined, each data point is assigned to the cluster of its nearest neighbor with higher density.

\paragraph{3) Saddle point identification} This step identifies the saddle points in the density field, which serve as border points between clusters. A saddle point, call it $i$, between clusters $c$ and $c'$ is the data point that has the following properties: \textit{a}) has a point $j$ in its k-nearest neighborhood belonging to cluster $c'$ and $i$ is the nearest neighbor of $j$  belonging to $c$, and \textit{b}) $i$ is the point with higher density belonging to $c$ or $c'$ for which property \textit{a} holds.

\paragraph{4) Density peak validation and hierarchical merging} In this final step, the algorithm validates potential clusters. The density at the center of each cluster, $\log(\rho_c)$, is compared with the density at the border with each neighboring cluster $\log(\rho_{cc'})$. The two are merged if the condition $$\log(\rho_c) - \log(\rho_{cc'}) < Z (\varepsilon_c + \varepsilon_{cc'})$$ is met. This comparison helps to determine whether a cluster is a true density peak or merely the result of a statistical fluctuation. The value of $Z$ controls the merging of the provisional clusters identified in step 2; a larger $Z$ value leads to more aggressive merging. The complete procedure is detailed in \cite{derrico2021adp}. Figure~\ref{fig:ADP_procedure} reports a pictorial representation of the full \ADP\ procedure. 

\begin{figure*}[t]
    \centering
    \includegraphics[width=1.0\textwidth]{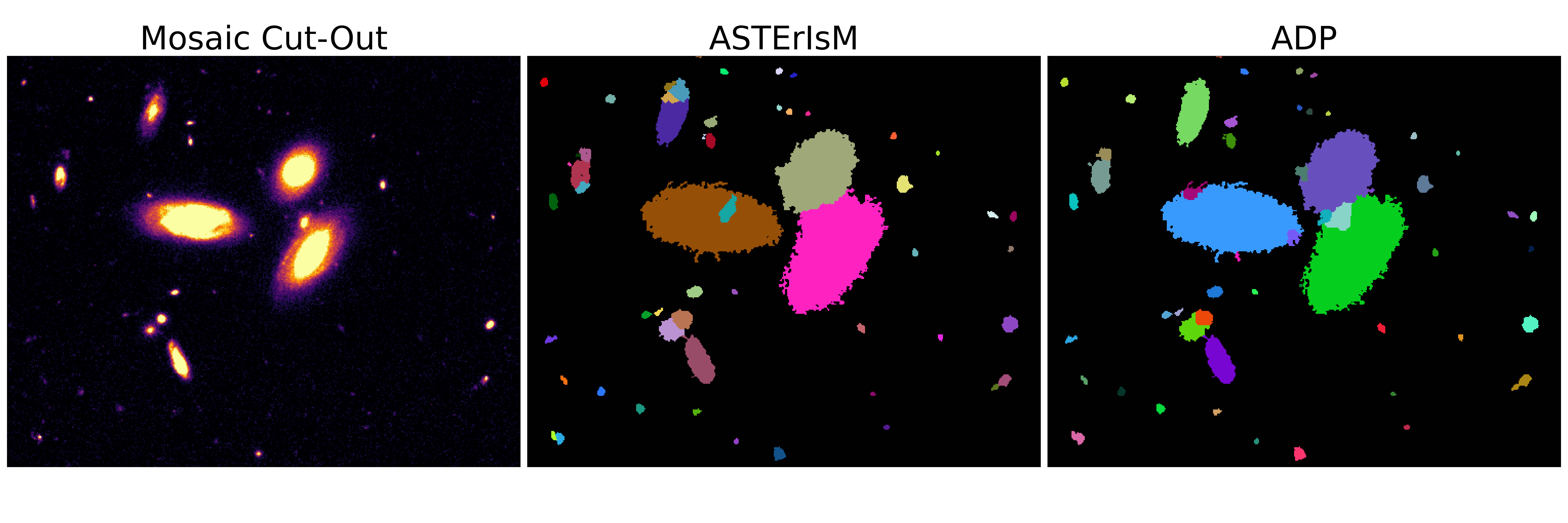
    }
    \caption{Comparison of image segmentation results. The left panel shows a 500×400 cutout from the EQ1. The middle panel presents the segmentation map of the same region obtained with \aste, while the right panel shows the result produced by \ADP. In both maps, the identified sources are color-coded.}
    \label{fig:Segmap_comparison}
\end{figure*}

To adapt \ADP\ for the deblending use case, two primary modifications were introduced. First, the individual data points are replaced with pixels from the image, such that the \textit{k-nearest} neighbors correspond to neighboring pixels within a square neighborhood of half-width $R$, defining a $(2R+1)\times(2R+1)$ pixel window. Second, the density values associated with each pixel are estimated from the mean flux of the pixels within this neighborhood, while the associated error is quantified through the standard deviation within the same region.

The adaptation of \ADP\ to astronomical images was accompanied by a significant effort to develop an efficient implementation optimized for High Performance Computing (HPC) infrastructures. The computational core is implemented in C and parallelized using OpenMP~\citep{openmp98}. The code is compiled into a shared object that can be used either from an executable or encapsulated into a Python module providing access to the compiled routines within astronomical analysis workflows.

The different stages of \ADP\ offer different degrees of parallelism. Wherever image pixels can be processed independently, a high degree of parallelism can be achieved, particularly in the following steps: \textit{i}) the initial conversion of luminosity values into density and error estimates; \textit{ii}) the identification of candidate cluster centers; and \textit{iii}) the identification of saddle points. However, the overall performance is ultimately constrained by inherently serial components, most notably the validation of density peaks in the final stage of the procedure. The validation of a given peak may depend on the finalized status or metrics of previously validated peaks, introducing a sequential dependency that limits the extent to which this stage can benefit from parallel execution. Despite its intrinsically serial nature, this stage accounts for at most approximately 5\% of the total execution time across the tested configurations.

A second level of parallelism is offered by inherent data-parallelism specific to the astronomical deblending. Because deblending follows a preliminary source-detection step, the full image can be decomposed into independent detection regions, each of which can be processed separately. The resulting patch-based decomposition therefore allows multiple detection regions to be deblended concurrently, providing an additional level of parallelism beyond that available within an individual patch and naturally enabling the workload to be distributed across larger computational resources.

Overall, this redesign provides several advantages for astronomical deblending with respect to \aste. \ADP\ follows a non-iterative procedure, while its patch-based formulation exposes parallelism both within and across detection regions. Furthermore, the combination of a density-based approach with a hierarchical clustering strategy reduces the sensitivity to variations in density among different sources. Finally, the identification of significant density structures is primarily controlled by the parameter $Z$, which directly regulates the conservativeness of the deblending procedure. Together with the low computational cost of \ADP, the limited number and direct interpretation of its free parameters allow the deblending configuration to be efficiently explored and fine-tuned for different datasets or particularly challenging source configurations.

These characteristics make the resulting implementation suitable for the processing of large astronomical images. The current implementation is optimized for CPU execution in shared-memory environments and has been applied to images from the EQ1 public data release, with dimensions up to 19200×19200 pixels.  As an illustrative example, Figure~\ref{fig:Segmap_comparison} shows a 500×400 cutout from the EQ1 data release alongside the corresponding segmentation maps produced by \ADP\ and \aste. While a quantitative comparison of their segmentation, morphological, photometric, and computational performance is presented in the following Sections, this example provides a visual impression of the segmentation output.
 
To enhance performance and scalability, future development will transition the codebase to a hybrid paradigm. First, GPU acceleration will be implemented via OpenMP offloading to ensure cross-platform portability and code readability~\citep[see for instance][]{Demeure_2023, Lepinza_porting, LACOPO2026101060}. Second, a distributed-memory framework via MPI will scale the application to multiple nodes, bypassing single-node memory constraints to enable the processing of massive datasets.

Following this latter direction, an MPI-based distributed-memory implementation of the \ADP, referred to as \DADP, has already been developed for its original clustering formulation as in~\cite{derrico2021adp}. This extension specifically targets tabular datasets, such as those arising in cosmological simulations, where \ADP\ serves as a substructure finder. The primary objective of this effort was to enable the algorithm to process datasets containing billions of data points. 

To achieve scalable performance, \DADP\ employs a distributed-memory implementation of a kd-tree for both domain decomposition and the \textit{k-nearest} neighbors search. Within the \ADP\ core procedure, efficiency is further enhanced through One-Sided Remote Memory Access (RMA) provided by the MPI framework~\citep{mpi41}. This feature enables data points to directly exchange information, such as density values and cluster assignments. Unlike the traditional Send/Receive scheme, one-sided access is more suitable in this context, as information from remote tasks is mostly accessed in read mode and does not require the simultaneous coordination of sender and receiver processes. A  detailed paper on this distributed architectural approach is currently in preparation~\citep{Tomba_in_prep}. 

Looking ahead, the deblending application represents a natural next key target for performance scaling, with future work focusing on the development of a distributed-memory version of the \ADP\ deblender. This effort will directly leverage optimizations and other solutions that have been developed for the case of tabular datasets in distributed memory environments.

\section{Validation and performance methodology}
\label{Validation_pipeline} 
A rigorous validation of a deblending algorithm ideally requires comparing the predicted segmentation against a reference GT. In astronomical imaging, however, such a reference is fundamentally unavailable. Segmentation maps provided by existing surveys are themselves the product of specific detection and deblending pipelines and therefore inherit the assumptions and limitations of the underlying algorithms. Consequently, using these products as GT may introduce systematic biases and does not provide an objective assessment of deblending performance. 

To overcome these limitations, we developed a modular validation framework specifically designed for astronomical deblending. The framework combines realistic image simulations, automatically generated GT segmentation, and label-invariant evaluation metrics, enabling a quantitative assessment of deblending performance under controlled conditions.

In addition to assessing the scientific performance of the algorithm, we evaluate its computational efficiency through strong- and weak-scaling experiments, together with an end-to-end runtime comparison with \aste\ on EQ1 observations. These benchmarks characterize the scalability of the OpenMP implementation as the computational resources and workload are varied, while the end-to-end comparison quantifies the practical computational advantage of \ADP\ under realistic processing conditions.

\subsection{Synthetic image generation}
\label{Synthetic_images}
The first component of the validation framework consists of a fully controllable simulation engine capable of generating realistic astronomical images with known source properties. The framework supports two complementary validation strategies. The first consists of controlled pairwise simulations, designed to isolate the influence of individual parameters on the deblending process. The second comprises synthetic multi-source images that provide a more realistic observational scenario while retaining a fully known GT.

Both point-like and extended objects are included in the simulations. Stars are modeled as two-dimensional Gaussian profiles, while galaxies are represented by single-component 2d Sérsic profiles. Individual sources are generated independently and combined to produce the final image. To reproduce \textit{Euclid} VIS-like~\citep{EuclidVIS} observing conditions, the simulated images incorporate a Gaussian background whose mean and standard deviation are estimated from the background maps of the EQ1 public data release. 

The final images are subsequently convolved with a Moffat point-spread function (PSF) characterized by a fixed full width at half maximum (FWHM) of 2 pixels and a shape parameter $\alpha = 3.5$. The Moffat profile is adopted as a simplified representation of the effective PSF, providing a well-defined spatial-resolution scale while retaining extended wings. The adopted value of $\alpha$ is chosen to provide a representative PSF shape and is not intended to reproduce the detailed \textit{Euclid} VIS PSF. The corresponding scale parameter $\gamma$, is derived from the adopted FWHM by inverting the standard Moffat relation: \begin{equation}
    \gamma = \frac{\mathrm{FWHM}}
    {2\sqrt{2^{1/\alpha}-1}}.
\end{equation} 

\subsubsection{Pairwise simulation}
\label{pairwise_sim}
The controlled pairwise simulations are designed to investigate the response of the deblending algorithms to the two main factors governing source confusion: the angular separation between neighboring objects and their relative brightness. Accordingly, the separation between the source centers and the flux ratio are systematically varied throughout the experiments. The remaining morphological parameters are fixed to representative mean values derived from the EQ1 catalogs. Although this choice does not capture the full diversity of observed source morphologies, it provides a controlled yet realistic benchmark that isolates the impact of blending while preserving the characteristic properties of \textit{Euclid} observations. Three classes of source pairs are considered throughout the analysis, namely star-star, star-galaxy, and galaxy-galaxy systems.

\subsubsection{Multi-source simulation}
\label{multisource_sim}
To complement the pairwise idealized experiments, the validation framework also includes synthetic multi-source images containing, on average, approximately thirty sources distributed over a 1500×1500 image size. Unlike the pairwise simulations, these images simultaneously contain isolated sources, partially blended systems, and more complex source configurations within a single realization. They therefore provide an intermediate validation stage between controlled pairwise experiments and real astronomical observations, combining a realistic spatial distribution of sources with the fundamental advantage of a completely known GT.

\subsubsection{From ideal to detection-limited GT}
\label{final_GT_creation}
For each simulated image, a corresponding GT segmentation map is generated using the importance-based pixel assignment strategy proposed by~\cite{haigh2021optimising}. Rather than assigning pixels solely according to the brightest contributing source, this method jointly considers both the statistical significance of each source at a given pixel and its relative contribution to the total pixel intensity. 

The resulting synthetic segmentation map is then restricted to the detected regions by masking it with a detection segmentation map produced by \se\ on the corresponding simulated image, using the same detection configuration adopted for the EQ1 VIS-processing as reported in~\cite{Euclid:2025wfi}. This removes pixels that would not be available to the deblending algorithm, ensuring that the evaluation is performed exclusively within the regions identified during the detection stage. Figure~\ref{fig:GT_creation_example} illustrates the construction of the reference GT segmentation for two representative validation scenarios: a galaxy--galaxy pair and a synthetic multi-source image. This procedure provides the reference segmentation adopted throughout the validation presented in the following Sections. 

\subsection{Pairwise evaluation metrics}
\label{Pairwise_metrics}
A second challenge in the validation of deblending algorithms concerns the choice of appropriate evaluation metrics. For the controlled pairwise simulations described in Section~\ref{pairwise_sim}, we adopt pairwise precision and pairwise recall, two label-invariant metrics commonly used to compare clustering solutions~\citep{Fowlkes_Mallows, Barnes2015}. 
\begin{figure}[t]
    \centering
    \includegraphics[width=0.5\textwidth]{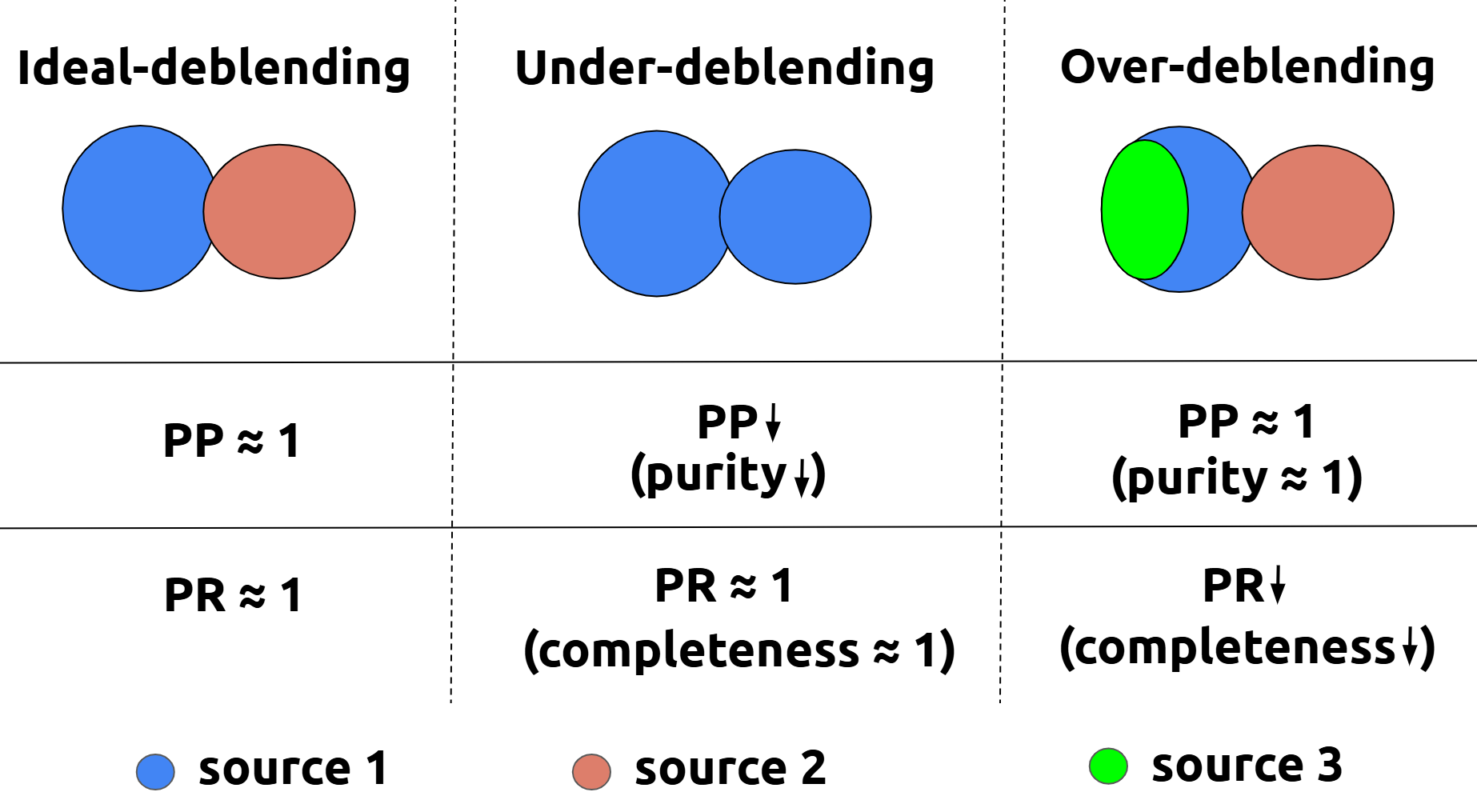}
    \caption{Illustration of the PP and PR behavior for an ideal segmentation (left), under-deblending (middle), and over-deblending (right). Under-deblending primarily reduces PP, whereas over-deblending primarily reduces PR.
    }
    \label{fig:PP_PR_interpretation}
\end{figure}
Unlike conventional segmentation measures, such as the pixel-wise confusion matrix or the Intersection over Union (IoU), these metrics are insensitive to label permutations and therefore do not require an explicit correspondence between the labels assigned to the predicted and GT segmentation maps. Although commonly used in the context of clustering analysis, they can be regarded as the pixel-pair analogues of catalog purity and completeness, respectively, providing an intuitive framework for evaluating the quality of a deblending solution from an astronomical perspective.

Formally let True Positives (TP) denote the number of pixel pairs assigned to the same source in both the predicted and GT segmentation, False Positives (FP) the number of pixel pairs grouped together only in the prediction, and False Negatives (FN) the number of pixel pairs grouped together only in the GT segmentation. Pairwise precision (PP) and Pairwise Recall (PR) are then defined as:
\begin{equation}
\begin{aligned}
\mathrm{PP} &= \frac{TP}{TP+FP},\\[0.2cm]
\mathrm{PR} &= \frac{TP}{TP+FN}.
\end{aligned}
\label{eq:pairwise_metrics}
\end{equation}

For the controlled pairwise simulations considered in this work, these metrics provide a direct characterization of the two main deblending failure modes. Since each image contains only two sources, under-deblending introduces pixel pairs that are incorrectly grouped together, reducing PP, whereas over-deblending separates pixel pairs belonging to the same source, reducing PR. Figure~\ref{fig:PP_PR_interpretation} illustrates these two limiting cases.

Although PP and PR quantify the quality of the recovered segmentation, they do not directly assess how differences in pixel assignment propagate to source photometry. We therefore complement this analysis by comparing the photometric measurements obtained from the \ADP\ and \aste\ segmentation maps. For each simulated realization, the catalogs produced by the two algorithms are cross-matched with the corresponding GT one, allowing both common recoveries and sources identified exclusively by either method to be characterized. 

\begin{figure}[t]
    \centering
    \includegraphics[width=0.5\textwidth]{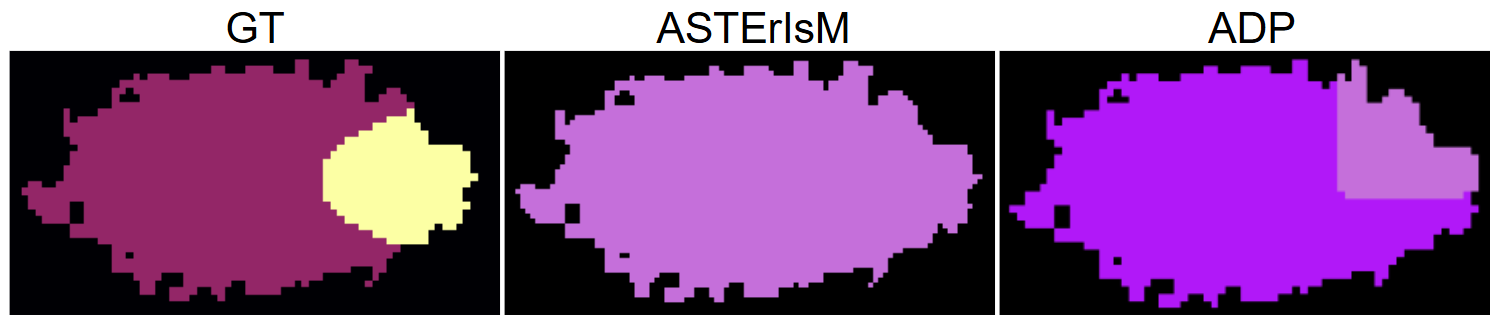}
    \caption{Illustration of the rationale behind the matching radius adopted for the pairwise photometric comparison. The left panel shows the GT segmentation, while the middle and right panels present the segmentation maps produced by \aste\ and \ADP, respectively. Although the segmentation recovered by \ADP\ is imperfect, the faint companion is partially detected, whereas it is missed by \aste. The resulting centroid displacement motivates the use of a larger matching radius to identify genuine detections in strongly blended configurations.}
    \label{fig:radius_motivation}
\end{figure}

For the pairwise simulations, a matching radius of 15 pixels is adopted. This relatively large value is motivated by the challenging blending configurations explored in these experiments, for which an imperfect reconstruction of the segmentation can produce substantial shifts in the recovered centroid even when the detection corresponds to the same simulated source. The adopted value was determined empirically by examining the number of GT matches as a function of matching distance and verifying that it reaches a plateau at approximately 15 pixels. Figure~\ref{fig:radius_motivation} illustrates one such example, in which \ADP\ successfully identifies the faint companion despite an imperfect reconstruction of its segmentation, whereas \aste\ fails to recover the source. 
For the common detections, the photometric consistency is assessed through the difference in \texttt{FLUX\_{ISOAREA}}, defined as the sum of the pixel fluxes assigned to each source in the corresponding segmentation map. 

\subsection{Multi-source evaluation metrics}
\label{multi_source_eval}
While PP and PR provide a direct interpretation of the two main deblending failure modes in controlled two-source experiments, this correspondence becomes less direct in multi-source images, where under- and over-deblending may occur simultaneously within the same realization. The validation therefore shifts from characterizing individual deblending failures to assessing the recovery of the source catalog as a whole. 

In this context, we valuate the ability of each deblending algorithm to recover the simulated source population and the positional consistency with respect to the corresponding GT catalog. The analysis is performed over an ensemble of 400 independent realizations, providing a statistically representative sample of different source configurations. 

For each realization, the catalog produced by the two deblending algorithm is cross-matched with the corresponding GT catalog using a greedy matching algorithm that pairs the nearest sources between the two catalogs, up to a maximum distance of n pixels. The resulting matches are first used to characterize the number of recovered sources as a function of the maximum matching distance. For the subset of sources recovered by both \ADP\ and \aste, the relative positional offsets between the two catalogs are subsequently measured in order to assess the consistency of the source centroids returned by the two deblending algorithms.

\subsection{EQ1 validation}
\label{EQ1_validation}
Following the validation on simulated datasets, \ADP\ was applied to real observations from the EQ1 public data release. Unlike the simulated experiments, where a reference GT segmentation is available, the objective of this analysis is not to assess the absolute accuracy of the deblending, but rather to compare the consistency of the source properties recovered by \ADP\ and \aste\ under realistic observational conditions.  

A common source detection catalog was first generated using \se\ and subsequently processed by both deblenders. The resulting \ADP\ and \aste\ catalogs were cross-matched according to their centroid positions using a maximum separation of two pixels. While the nominal \textit{Euclid}-VIS PSF FWHM provides a characteristic resolution scale of $0.1"$~\citep{EuclidVIS}, corresponding to one pixel, adopting this value as the matching threshold would impose a particularly restrictive criterion on the centroids independently recovered by the two deblenders. 

The robustness of the matching criterion was therefore assessed by progressively increasing the matching radius, using one random picked observation. A one-pixel threshold yields 90.007 matched sources, while increasing the radius to two pixels raises this number to 94.622, corresponding to an increase of approximately 5\%. Increasing the threshold further to three pixels results in 97.593 matches, an additional increase of only 3\%, with progressively smaller variations at larger radii. Moreover, the morphological and photometric distributions remain stable beyond $2-3$ pixels. A two-pixel radius was therefore adopted as a compromise between recovering genuine counterparts and limiting potentially ambiguous associations between nearby deblended components.

The morphological analysis considered the \texttt{ISOAREA}, \texttt{ELLIPTICITY}, and \texttt{POSITION\_ANGLE}. Here, \texttt{ISOAREA} corresponds to the number of pixels assigned to each source in the segmentation map providing a direct measure of the recovered source extent. The \texttt{ELLIPTICITY} and \texttt{POSITION\_ANGLE} are instead derived from the covariance matrix of the pixels assigned to each source, with its eigenvalues determining the extension along the main axes and its eigenvectors defining their orientation. 

In addition, the photometric consistency of the two deblenders is assessed by comparing the magnitudes measured from their respective segmentation maps using \texttt{A-PHOT}~\citep{Merlin2019}. For each source, the photometric aperture used by \texttt{A-PHOT} are defined according to the previously described morphological parameters estimated from the corresponding segmentation maps.

\subsection{Performance evaluation}
\label{strong_and_weak_methodology}
To characterize the computational performance of \ADP, both strong- and weak-scaling experiments were carried out on a shared-memory architecture using the OpenMP implementation of the algorithm. For each configuration, the execution was repeated ten times on the same input data to mitigate the impact of runtime fluctuations and operating-system noise. In addition to the total execution time, the runtime of the main computational stages, namely the density estimation, the clustering, the segmentation map generation, the source property computation and the catalog export, were measured independently to identify the dominant computational components and assess their scalability.

In the strong-scaling experiments, a fixed EQ1 detection segmentation map was processed while increasing the number of OpenMP threads from 1 to 32. This benchmark evaluates the reduction in execution time achievable for a fixed computational workload and provides a measure of the parallel efficiency of the implementation.

For the weak-scaling experiments, the computational workload assigned to each OpenMP thread was kept approximately constant while increasing the number of threads. Since \ADP\ operates independently on the detection patches identified by \se, each detection patch was treated as an independent computational task, with its \texttt{ISOAREA} adopted as a proxy for the associated workload. 

Detection patches were sorted according to their \texttt{ISOAREA} and distributed among the available threads using a greedy load-balancing strategy, in which each successive patch was assigned to the thread with the lowest accumulated workload. This procedure generated groups of detection patches with approximately equal total \texttt{ISOAREA} while preserving the original segmentation map labels. For each thread configuration, the corresponding detection segmentation map was constructed by retaining only the selected detection patches, such that the total computational workload increased approximately proportionally to the number of OpenMP threads while maintaining a nearly constant workload per thread.

For the strong-scaling experiments, the speedup $S(n)$ and parallel efficiency $E_S(n)$ are defined as:
\begin{equation}
S(n)=\frac{T(1)}{T(n)}, \qquad
E_S(n)=\frac{S(n)}{n}=\frac{T(1)}{n\,T(n)},
\label{eq:parallel_efficiency_strong}
\end{equation}
where $T(n)$ denotes the execution time obtained using $n$ OpenMP threads.
For the weak-scaling experiments, where the workload per compute unit is kept constant, the two definitions coincide and are given by 
\begin{equation}
E_W(n)=\frac{T(1)}{T(n)} \equiv S(n) .
\label{eq:parallel_efficiency_weak}
\end{equation}

\subsection{End-to-end runtime evaluation}
\label{end-to-end-methodology}
While the scaling experiments  characterize the parallel behavior of the OpenMP implementation, they do not directly quantify the computational cost of applying \ADP\ to real astronomical observations. To assess the time-to-solution of the deblending process, an end-to-end runtime comparison between \ADP\ and \aste\ was performed using a representative sample of ten images from the EQ1 public data release. 
\begin{figure*}[t]
    \centering
        \includegraphics[width=1.0\textwidth]{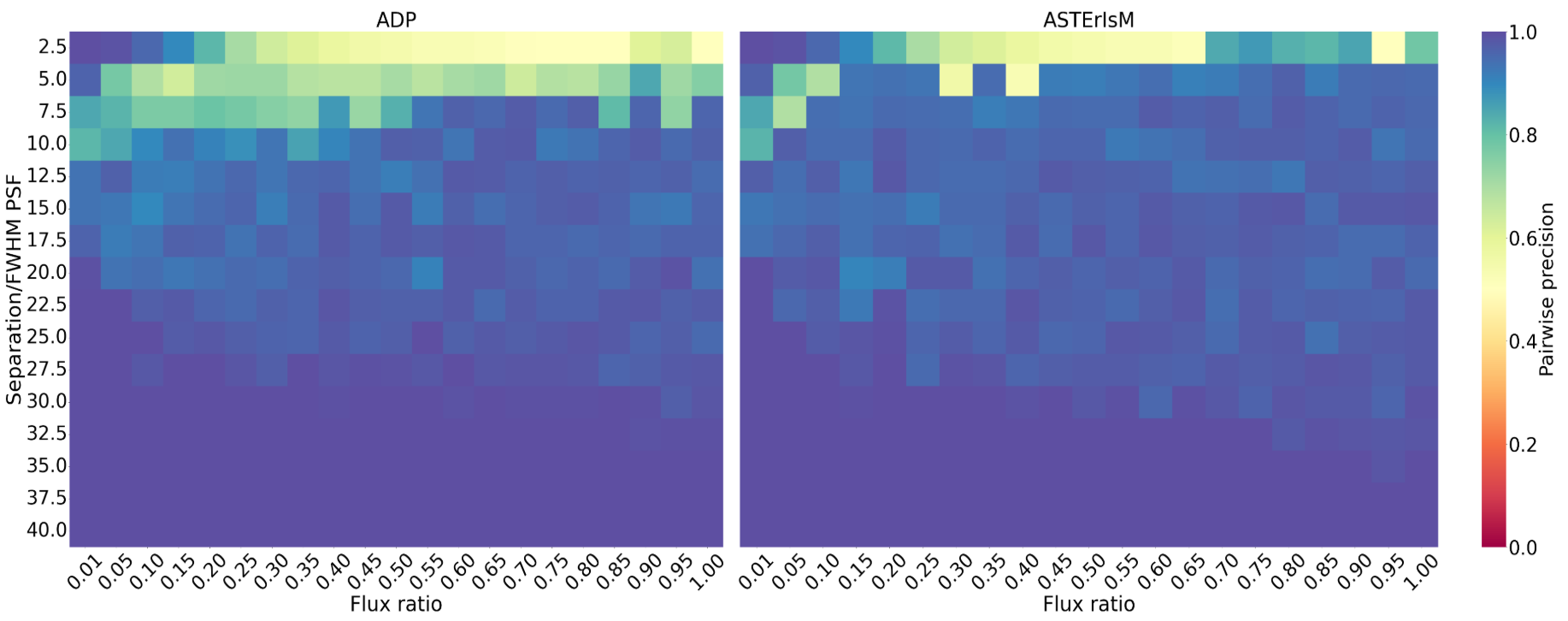}
    \caption{
    PP as a function of source separation and flux ratio for the galaxy-galaxy pairwise simulations. The left panel shows the results obtained with \ADP, while the right panel shows the results obtained with \aste.
    }
    \label{fig:pp_comparison}
\end{figure*}
\begin{figure*}[h!]
    \centering
        \includegraphics[width=1.0\textwidth]{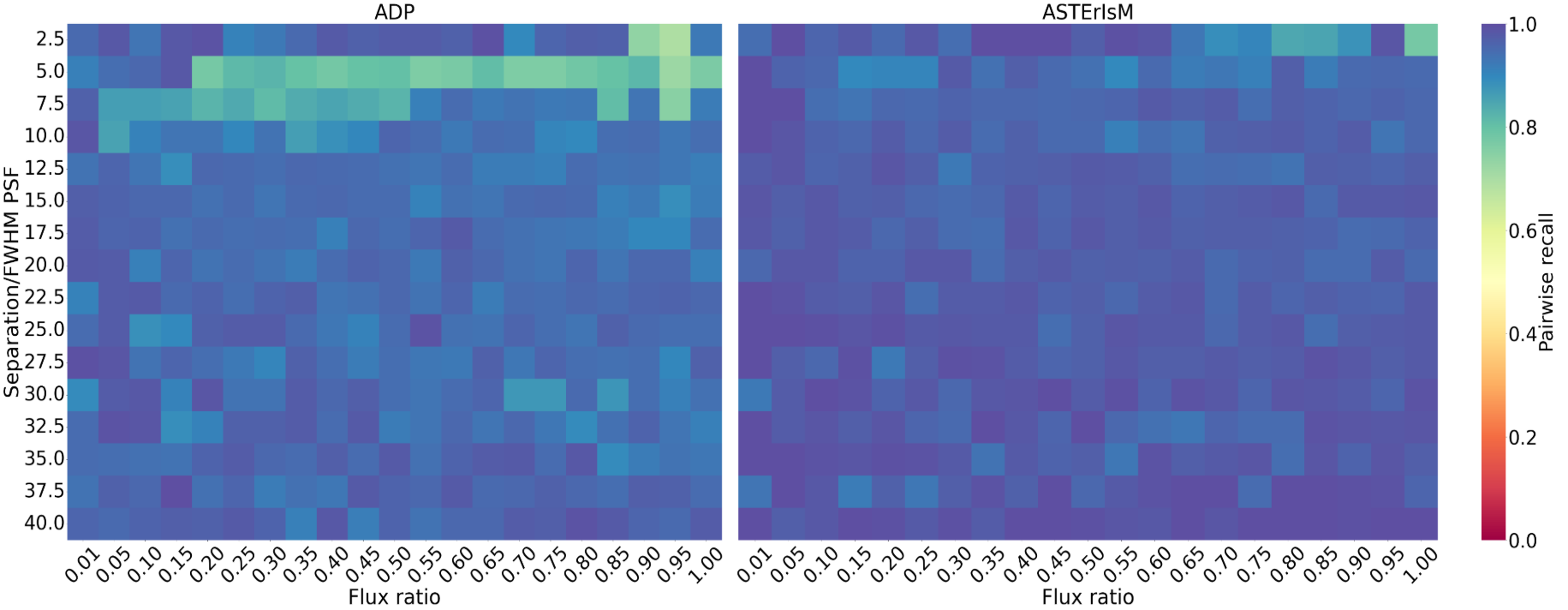}
    \caption{
    PR as a function of source separation and flux ratio for the galaxy-galaxy pairwise simulations. The left panel shows the results obtained with \ADP, while the right panel shows the results obtained with \aste.
    }
    \label{fig:pr_comparison}
\end{figure*}

The benchmark was performed on PLEIADI~\citep{bertocco2019inaf, taffoni2020chipp} platform, using a compute node equipped with two Intel Xeon E5-2697 v4 CPUs at 2.30~GHz, providing 36 physical cores (72 hardware threads) and 256~GB of memory. Each execution was restricted to 12 OpenMP threads and 64~GB of memory. For each image, both algorithms were executed starting from the same detection segmentation map and using identical computational resources. The total wall-clock time required to generate the final segmentation map and source catalog was recorded. This benchmark complements the scaling analysis by directly quantifying the computational cost of the two deblending approaches under realistic operating conditions.

\section{Results}
\label{Results}
In this Section, we evaluate the performance of \ADP\ through a progressive validation strategy described in Section~\ref{Validation_pipeline}, using \aste\ as the benchmark deblending algorithm adopted throughout this work. We first consider controlled pairwise simulations, described in Section~\ref{pairwise_sim}, to investigate the deblending behavior under well-defined conditions. The analysis is then extended to synthetic multi-source simulations introduced in Section~\ref{multisource_sim}, where the focus shifts from individual blending configuration to the recovery of the source catalog as a whole. Subsequently, the comparison is performed on real observations from the EQ1 public data release following the methodology described in Section~\ref{EQ1_validation}, allowing the consistency of the morphological and photometric properties recovered by the two deblenders to be assessed. Finally, the computational performance is evaluated through the strong- and weak-scaling experiments described in Section~\ref{strong_and_weak_methodology}, together with the end-to-end runtime comparison between \ADP\ and \aste\ presented in Section~\ref{end-to-end-methodology}.

\subsection{Controlled pairwise simulations}
\label{controlled_pairwise_gal_gal}
\begin{figure*}[h!]
    \centering
    \includegraphics[width=0.80\textwidth]{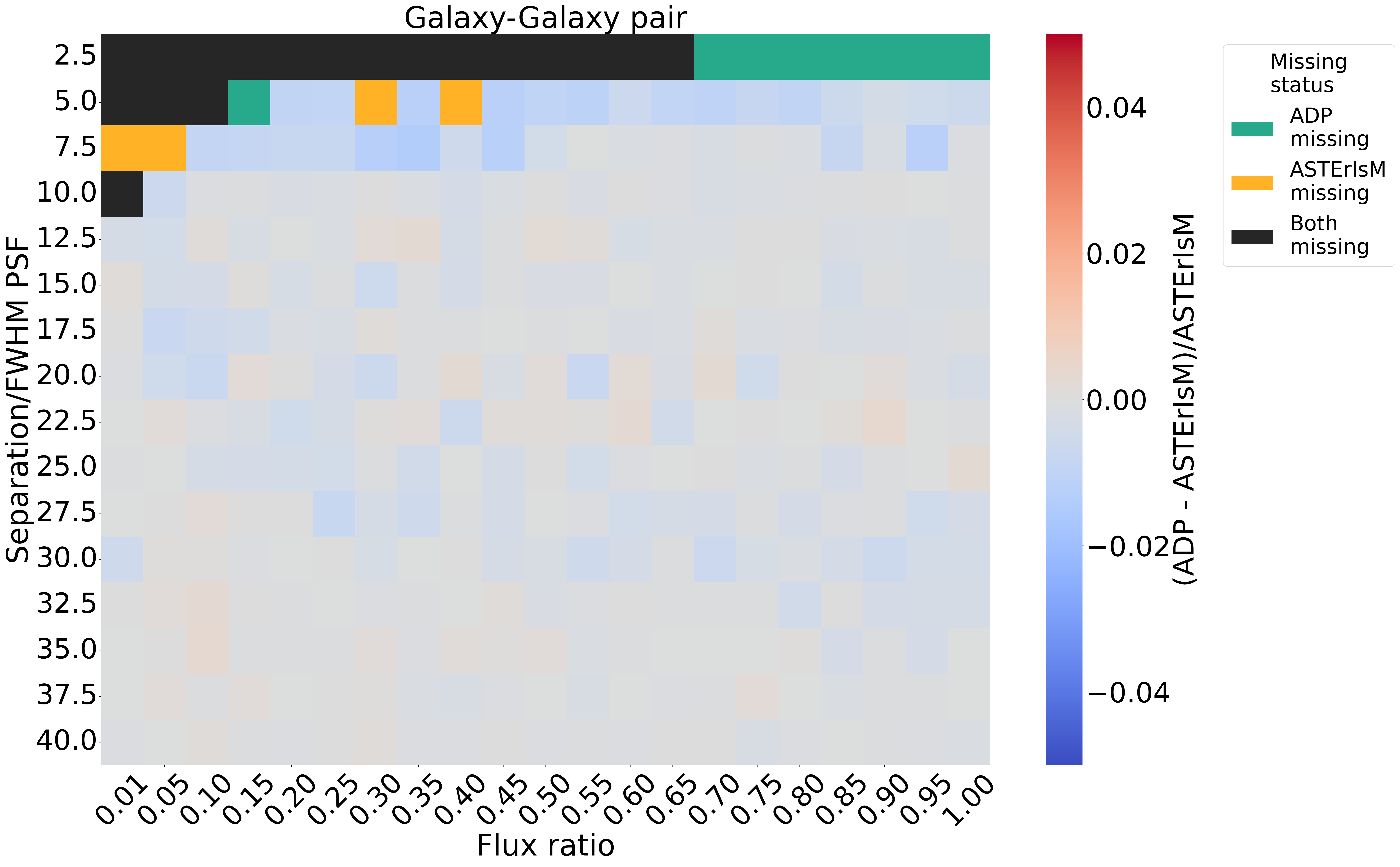}
    \caption{Residual of \texttt{FLUX\_{ISOAREA}} measured by \ADP\ and \aste\ for the matched sources, as a function of source separation and flux ratio for the galaxy-galaxy pairwise simulations. Dedicated colors highlight the configurations in which one source is missed by \ADP, one source is missed by \aste, or one source is missed by both.}
    \label{fig:pair_phot_comparison}
\end{figure*}

We begin by analyzing the performance of \ADP\ on the controlled pairwise simulations, comparing its behavior with \aste\ across the explored source separations and flux ratios. The results are examined separately for star-star, star-galaxy, and galaxy-galaxy systems, considering both the quality of the recovered segmentation through PP and PR and the corresponding photometric measurements.

Figures~\ref{fig:pp_comparison} and~\ref{fig:pr_comparison} present the PP and PR~\eqref{eq:pairwise_metrics} measured for the galaxy-galaxy configuration as a function of the source separation, expressed in units of the Euclid VIS-like PSF FWHM (2 pixels), and the source flux ratio. Overall, the two deblenders exhibit a similar depende of both metrics on the separation and flux ratio, indicating a comparable deblending performance across the explored parameter space. \aste\ reaches marginally higher PP and PR values in some regions, although the differences between the two methods remain small and become appreciable mainly at source separations below approximately $5-10$ pixels. For extended sources, such separations correspond to strongly overlapping configurations, where the deblending task becomes intrinsically challenging and small differences in the adopted segmentation strategy lead to measurable discrepancies.

This overall consistency is also reflected in the photometric comparison shown in Figure~\ref{fig:pair_phot_comparison}, where the \texttt{FLUX\_{ISOAREA}} measurements obtained from the segmentation maps produced by \ADP\ and \aste\ are compared. The Figure also indicates, using dedicated colors distinct from the flux-residual colormap, the configurations in which one source is missed by \ADP, one source is missed by \aste, or one source is missed by both. For these cases, no flux residual is reported, since a one-to-one photometric comparison is not available. The recovered fluxes show excellent agreement, with residual typically below 1\% and reaching approximately $5-7\%$ only in the most challenging configurations. No systematic offset between the two methods is observed, with either deblender yielding the larger flux depending on the specific blending configuration. This further indicates that the differences observed in PP and PR translate into only limited variations in the recovered source photometry. Equivalent conclusions are obtained for the star--star and star--galaxy configurations, whose results are presented in~\ref{Controlled_pairwise_additional}.

\subsection{Controlled multi-source simulations}
\begin{figure*}[h!]
\centering
\includegraphics[width=0.95\textwidth]{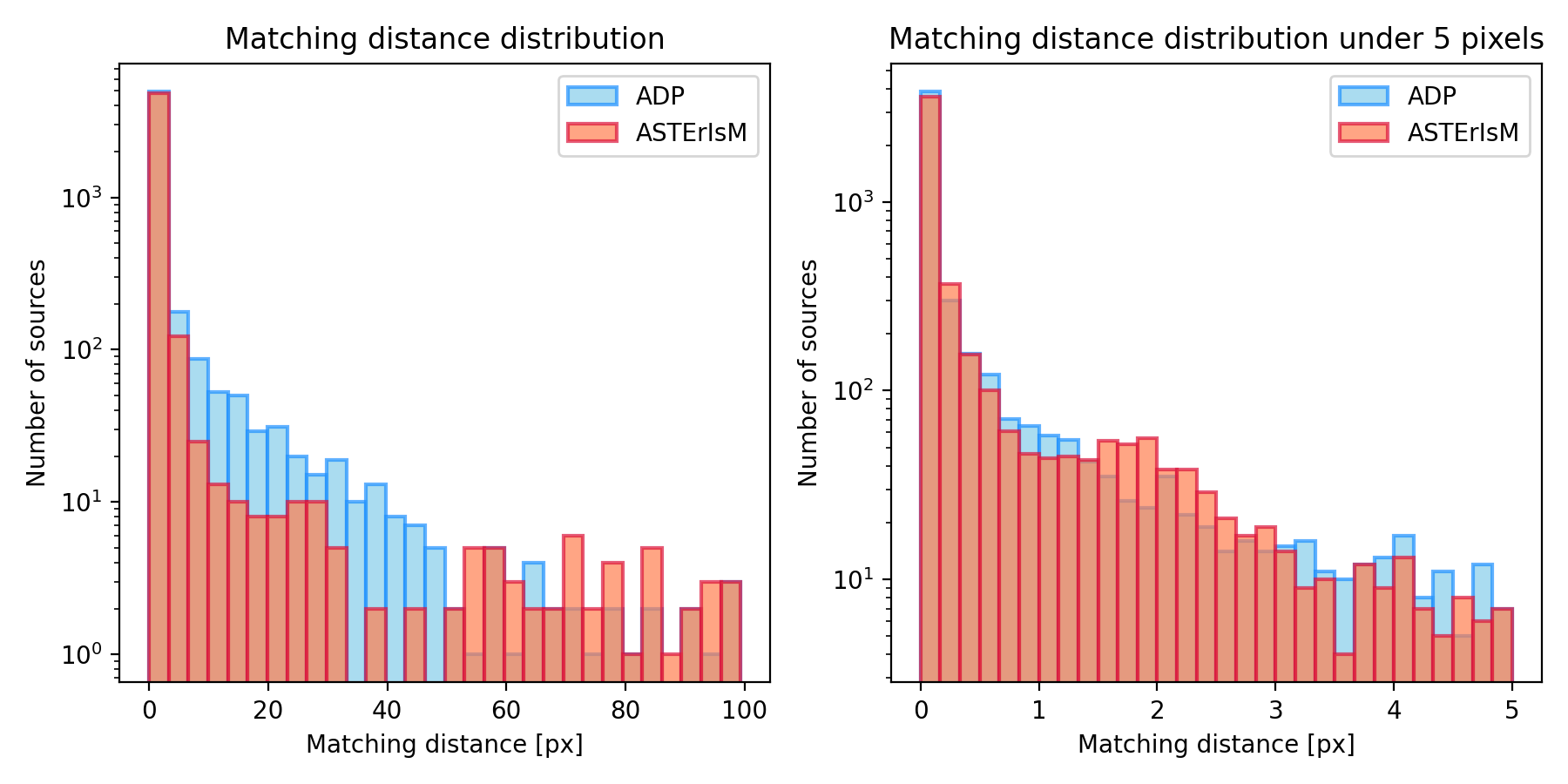}
\caption{Distribution of the matching distances between the ground GT and \aste, and between the GT and \ADP. The left panel shows the full distribution, while the right panel provides a detailed view restricted to matching distances of up to five pixels.}
\label{fig:matching_distance_distribution}
\end{figure*}
\begin{figure*}[h!]
\centering
\includegraphics[width=0.95\textwidth]{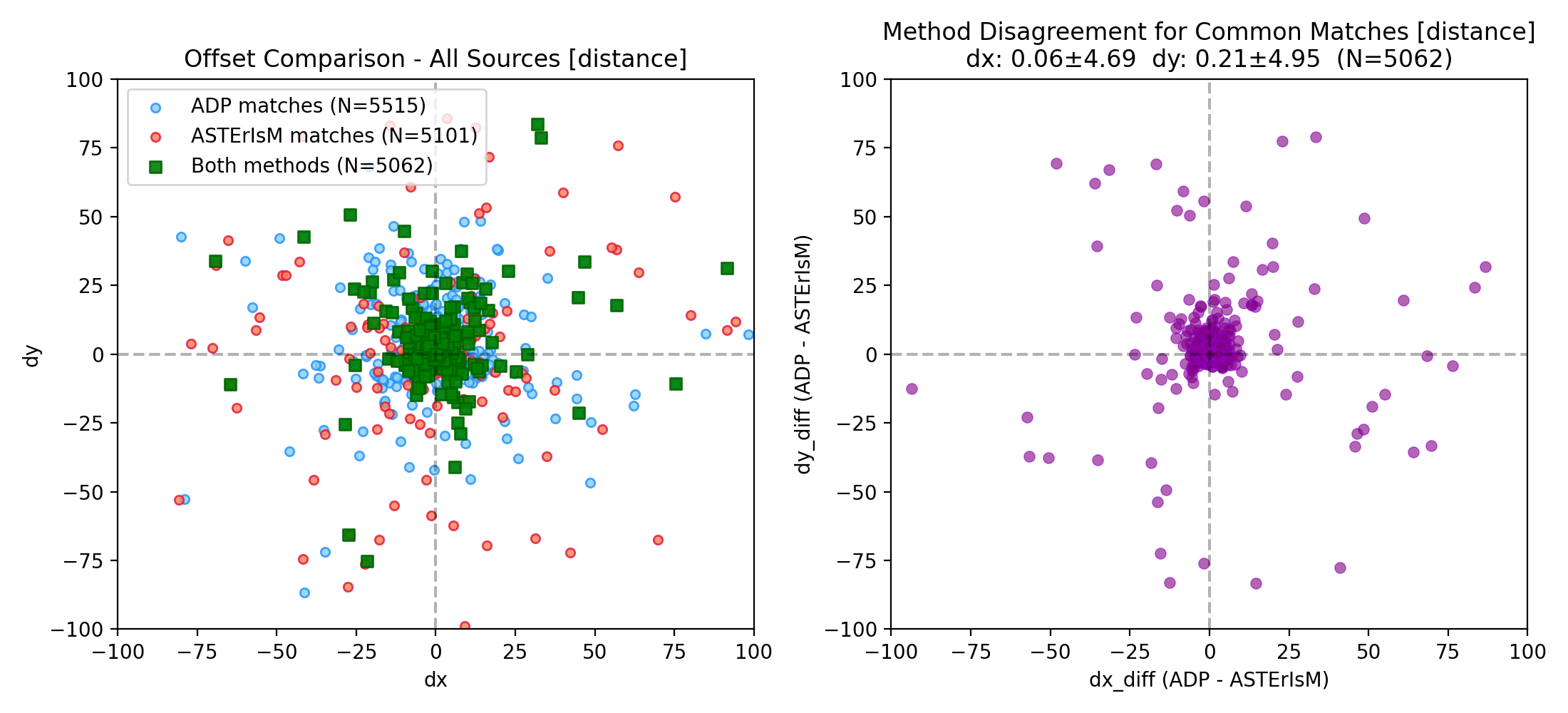}
\caption{Matching offset distribution. The offset is computed as the difference in the x and y coordinates between the centers of mass of matching sources. The green squares represent the sources matched with the GT by both \aste\ and \ADP. In the right panel, the offsets between common matches are shown as purple dots. As indicated in the subtitle, the mean offsets in both directions are close to zero and exhibit comparable variances, indicating no apparent systematic geometric bias between the two methods.}
\label{fig:matching_offsets}
\end{figure*}

The comparison is now extended to the synthetic multi-source images, where the performance of the two deblenders is assessed in terms of source recovery and positional accuracy over the ensemble of 400 simulated realizations. Figures~\ref{fig:matching_distance_distribution} and~\ref{fig:matching_offsets} summarize the resulting catalog-level comparison.

Figure~\ref{fig:matching_distance_distribution} shows the number of sources successfully matched to GT catalog as a function of the matching distance. \ADP\ recovers a larger number of GT sources than \aste\ across almost the entire range of matching distances. The excess of matches at small separations suggest an improved source recovery rate, particularly in crowded regions and for faint or strongly blended objects. Additional matches are also present at larger matching distances. Visual inspection suggests that these cases are primarily associated with sources embedded in the diffuse light of nearby bright objects, where the segmentation boundaries become difficult to determine accurately. In these situations, the segmentation maps produced by \ADP\ may differ significantly from the idealized GT segmentation, leading to larger centroid offsets and consequently larger matching distances. Nevertheless, these detections correspond to genuine sources that are correctly recovered by \ADP, indicating that the increased matching distance reflects differences in the inferred segmentation geometry rather than a failure to identify the underlying source. 

The positional offsets with respect to the GT catalog are shown in the left panel of Figure~\ref{fig:matching_offsets}. Overall, \ADP\ recovers 5515 GT sources, compared with 5101 for \aste, corresponding to approximately 8\% more matched sources. Despite this difference in source recovery, the two deblenders exhibit broadly comparable distributions of positional offsets with respect to the GT, indicating a similar level of positional accuracy.

Of the recovered sources, 5062 are common to both deblenders. The right panel of Figure~\ref{fig:matching_offsets} compares the reconstructed source positions obtained by \ADP\ and \aste\ for this common sample. The positional differences in both coordinates are centered around zero, with the two methods agreeing at the subpixel level and showing no evidence of a systematic positional offset. Overall, these results indicate that \ADP\ achieves a higher source recovery rate than \aste\ without compromising the positional accuracy of the recovered sources.

\subsection{Euclid Q1 comparison}
The comparison on real EQ1 public data release focuses on the consistency of the morphological and photometric properties recovered by \ADP\ and \aste\ in the absence of a reference GT. As described in Section~\ref{EQ1_validation}, the analysis is performed on sources successfully cross-matched between the two catalogs within the adopted matching radius of two pixels.

\begin{figure}[t]
    \centering
    \includegraphics[width=0.5\textwidth]{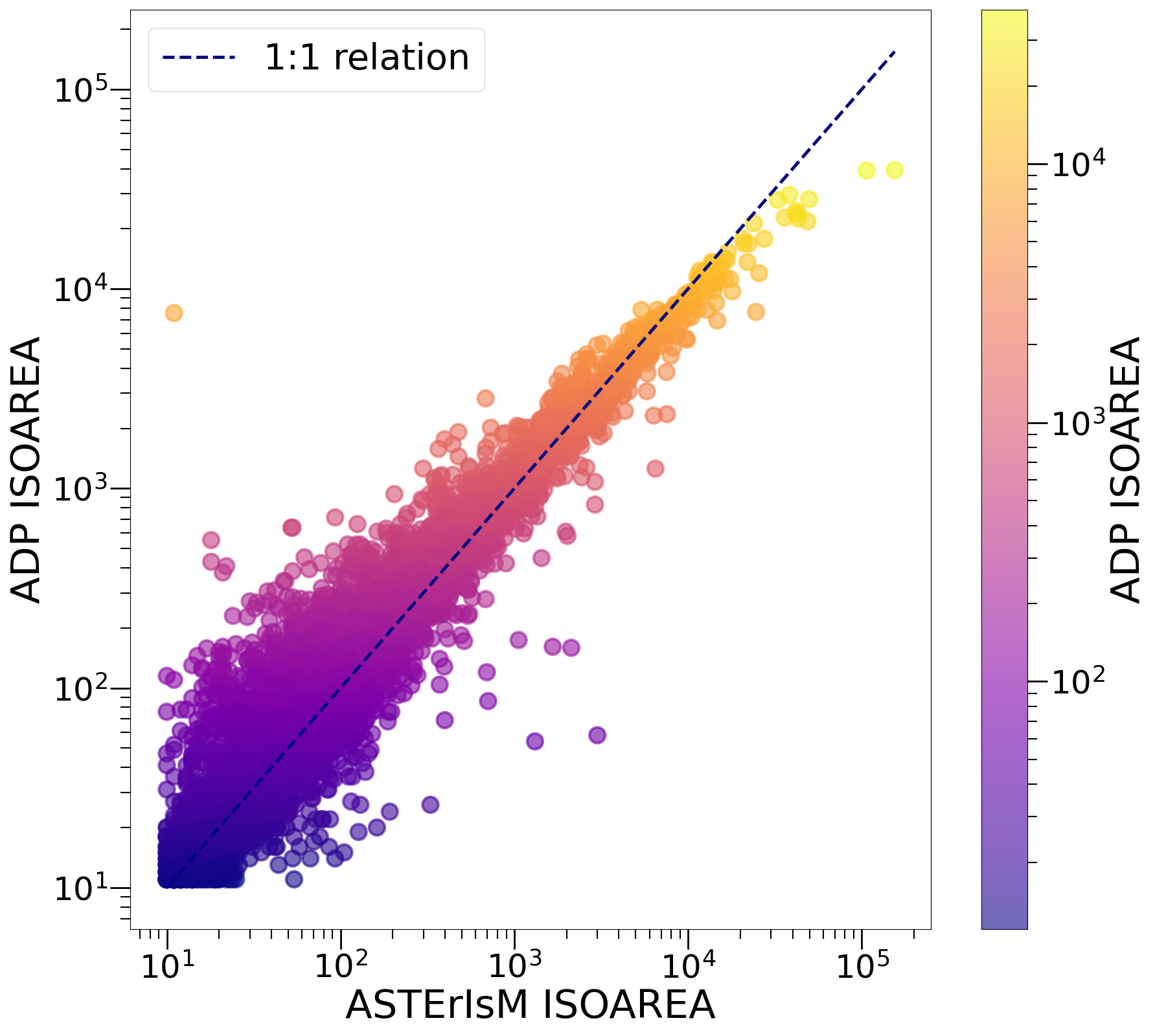}
    \caption{
    Comparison of the \texttt{ISOAREA} measurements obtained from the \ADP\ and \aste\ deblended catalogs for sources successfully cross-matched between the two datasets. Points are color-coded according to the \ADP\ \texttt{ISOAREA}, highlighting different source-size regimes. The dashed line indicates the one-to-one relation.
    }
    \label{fig:q1_isoarea}
\end{figure}

\begin{figure}[h!]
    \centering
    \includegraphics[width=0.5\textwidth]{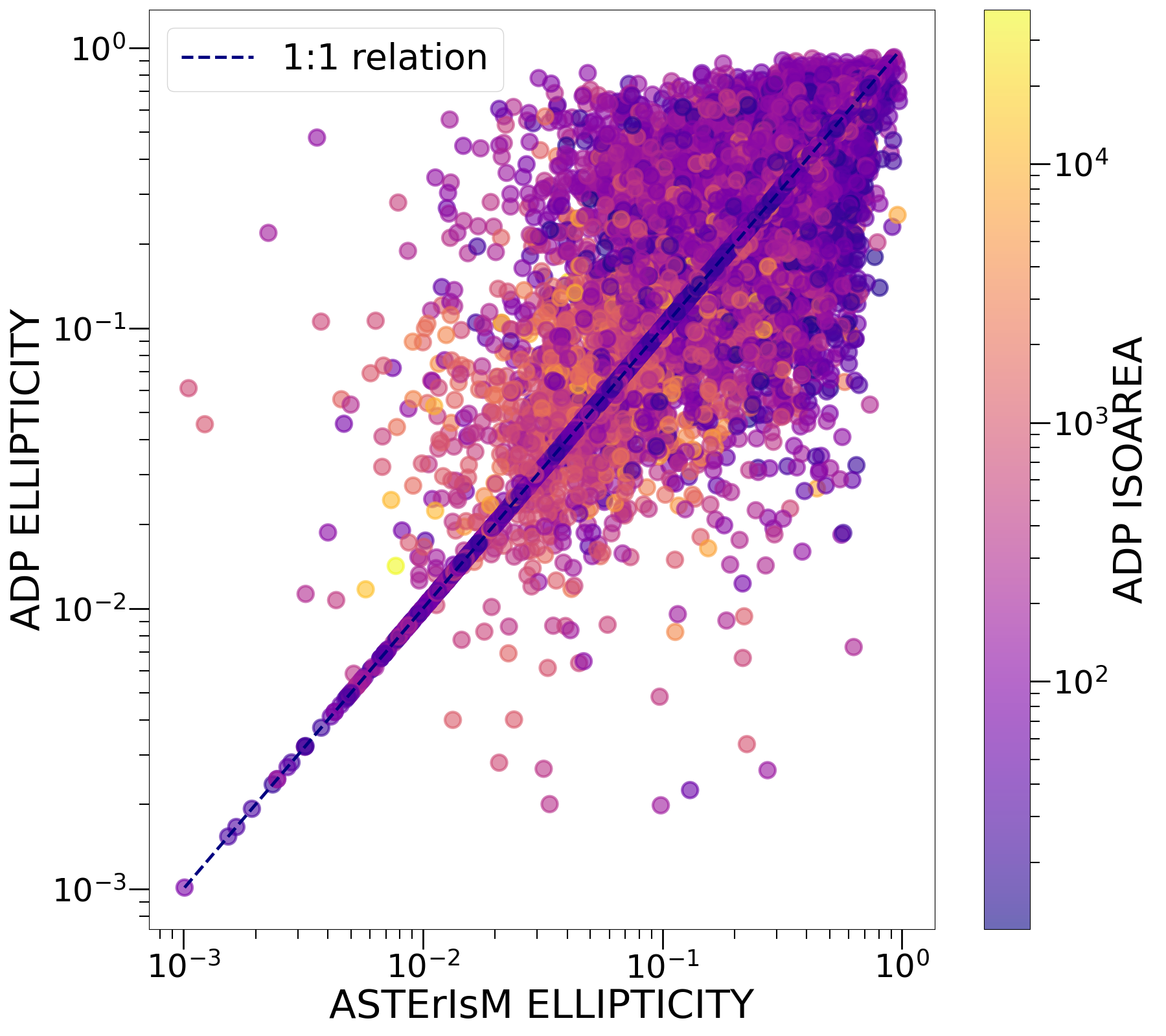}
    \caption{
    Comparison of the \texttt{ELLIPTICITY} measurements obtained from the \ADP\ and \aste\ deblended catalogs for sources successfully cross-matched between the two datasets. Points are color-coded according to the \ADP\ \texttt{ISOAREA}, highlighting different source-size regimes. The dashed line indicates the one-to-one relation.
    }
    \label{fig:q1_ellipticity}
\end{figure}

\begin{figure}[h!]
    \centering
    \includegraphics[width=0.5\textwidth]{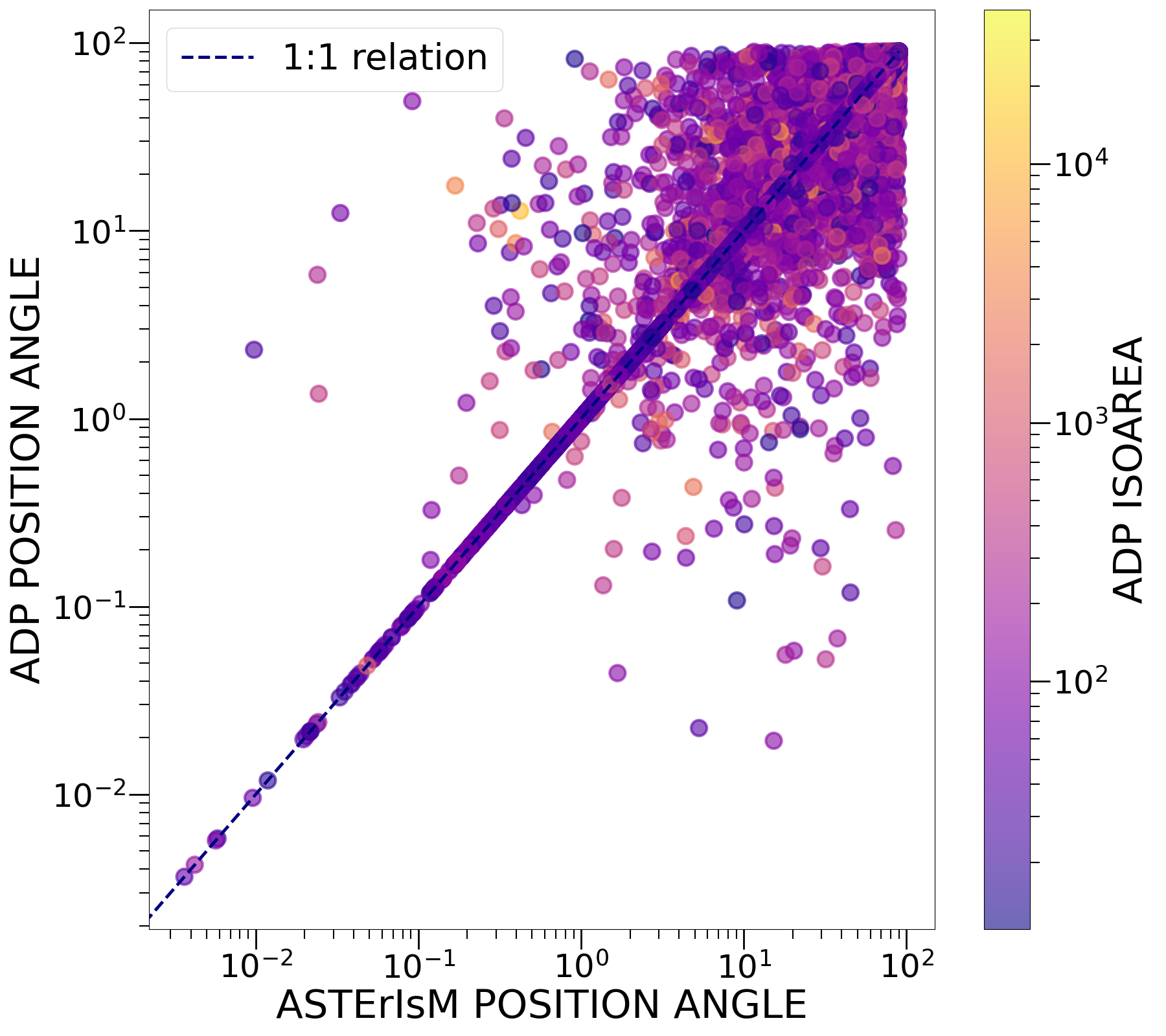}
    \caption{
    Comparison of the \texttt{POSITION\_ANGLE} measurements obtained from the \ADP\ and \aste\ deblended catalogs for sources successfully cross-matched between the two datasets. Points are color-coded according to the \ADP\ \texttt{ISOAREA}, highlighting different source-size regimes. The dashed line indicates the one-to-one relation.
    }
    \label{fig:q1_pos_angle}
\end{figure}

Figures~\ref{fig:q1_isoarea}, \ref{fig:q1_ellipticity}, and \ref{fig:q1_pos_angle} compare the morphological properties derived from the segmentation maps produced by the two deblenders, namely \texttt{ISOAREA}, \texttt{ELLIPTICITY}, and \texttt{POSITION\_ANGLE}. In each case, the values recovered by \ADP\ are compared with the corresponding \aste\ measurements, with sources color-coded according to their \ADP\ \texttt{ISOAREA}. This representation allows the agreement between the two methods to be examined as a function of the recovered source extent.

The \texttt{ISOAREA} measurements, shown in Figure~\ref{fig:q1_isoarea}, exhibit a strong overall agreement between the two deblenders across the full range of source extents. The scatter increases toward smaller sources, as expected given the discrete nature of the segmentation maps: for compact objects, differences in the assignment of only a few pixels can represent a significant fraction of the total segmentation area. Conversely, the agreement becomes increasingly tight for larger sources, with no clear systematic offset between the two methods.

The comparison of \texttt{ELLIPTICITY}, and \texttt{POSITION\_ANGLE}, presented respectively in Figure~\ref{fig:q1_ellipticity}, and \ref{fig:q1_pos_angle}, also shows a high degree of consistency between the two deblenders. The \ADP\ \texttt{ISOAREA} color coding reveals that sources with smaller areas lie predominantly along the one-to-one relation, while the scatter gradually increases toward larger and more elongated objects. This behavior is consistent with the definition of both quantities from the covariance matrix of the segmented pixels. Differences in pixel assignment can modify the inferred main axes of the source, affecting both its ellipticity and orientation and consequently increasing the scatter between the two deblenders.
\begin{figure}[t]
    \centering
    \includegraphics[width=0.5\textwidth]{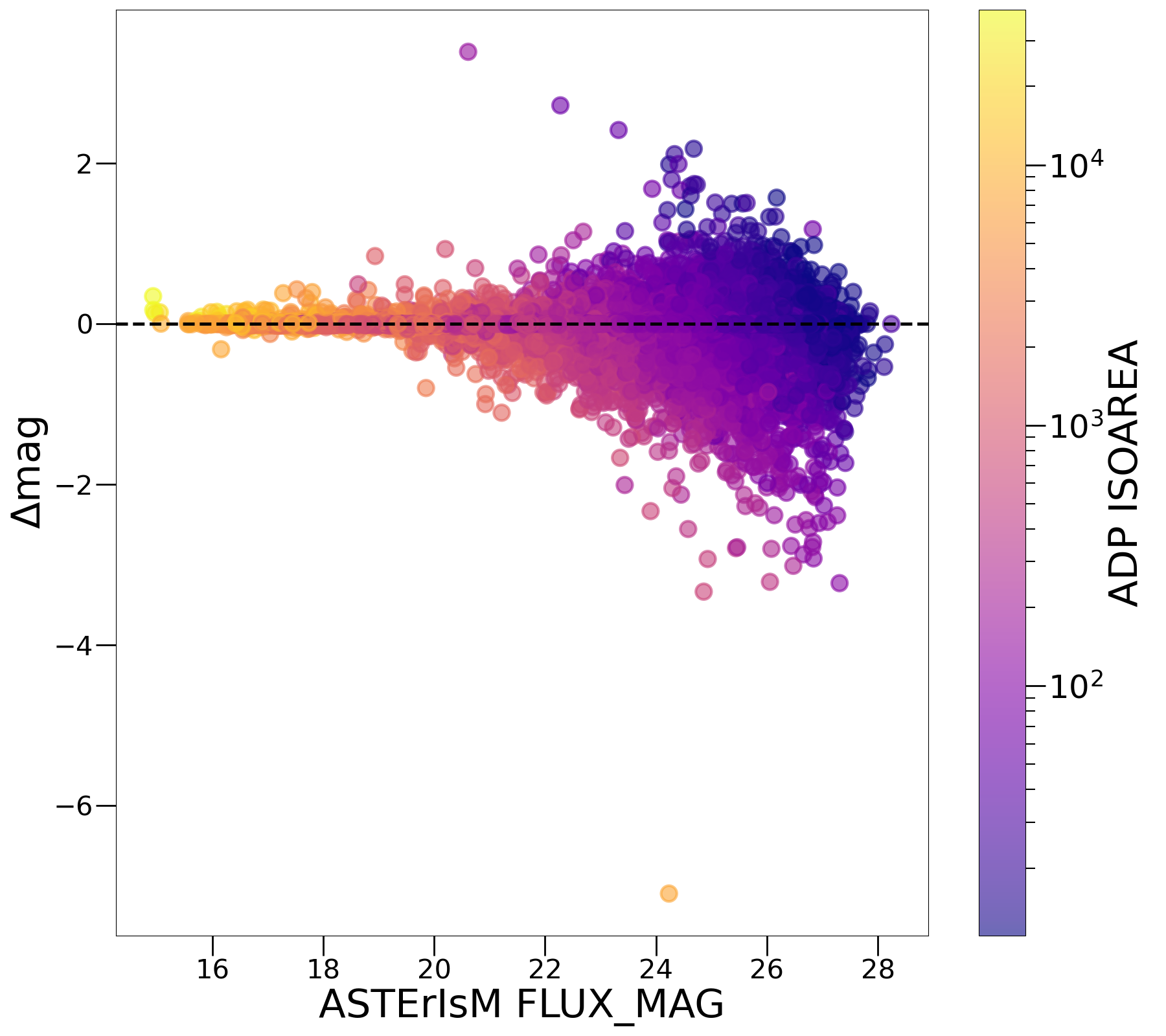}
    \caption{
    Magnitude difference, $\Delta\mathrm{mag}$, between the \ADP\ and \aste\ measurements as a function of \aste\ magnitude for the cross-matched sources in the EQ1 dataset. The points are colour-coded according to the \ADP\ \texttt{ISOAREA}. 
    }
    \label{fig:q1_photometry}
\end{figure}

The photometric comparison is presented in Figure~\ref{fig:q1_photometry}, which shows the magnitude difference, $\Delta\mathrm{mag}$, between the measurements obtained with \ADP\ and \aste\ as a function of the \aste\ source magnitude, with the points color-coded according to the \ADP\ \texttt{ISOAREA}. The two deblenders exhibit a strong overall photometric agreement, with the bulk of the distribution concentrated around $\Delta\mathrm{mag}=0$ across the full magnitude range. As expected, the scatter gradually increases toward fainter magnitudes, reflecting the lower S/N of these objects and the increased sensitivity of the recovered flux to small differences in the reconstructed segmentation. The largest photometric differences are predominantly associated with compact sources, while extended objects remain tightly clustered around zero. 

A mild asymmetry is visible for the smallest and faintest sources, for which \ADP\ tends to recover slightly brighter magnitudes than \aste, suggesting that differences in the reconstructed segmentation result in a somewhat larger fraction of the source flux being recovered by \ADP. Interestingly, this behavior is qualitatively consistent with the photometric flux deficit reported for EQ1 observations with respect to Gaia~\citep{Euclid:2025wfi}, although a dedicated validation against external photometric references will be required to assess whether the additional recovered flux corresponds to an improvement in the absolute photometric accuracy.

\subsection{Strong scaling}
\begin{figure}[t]
    \centering
    \includegraphics[width=0.5\textwidth]{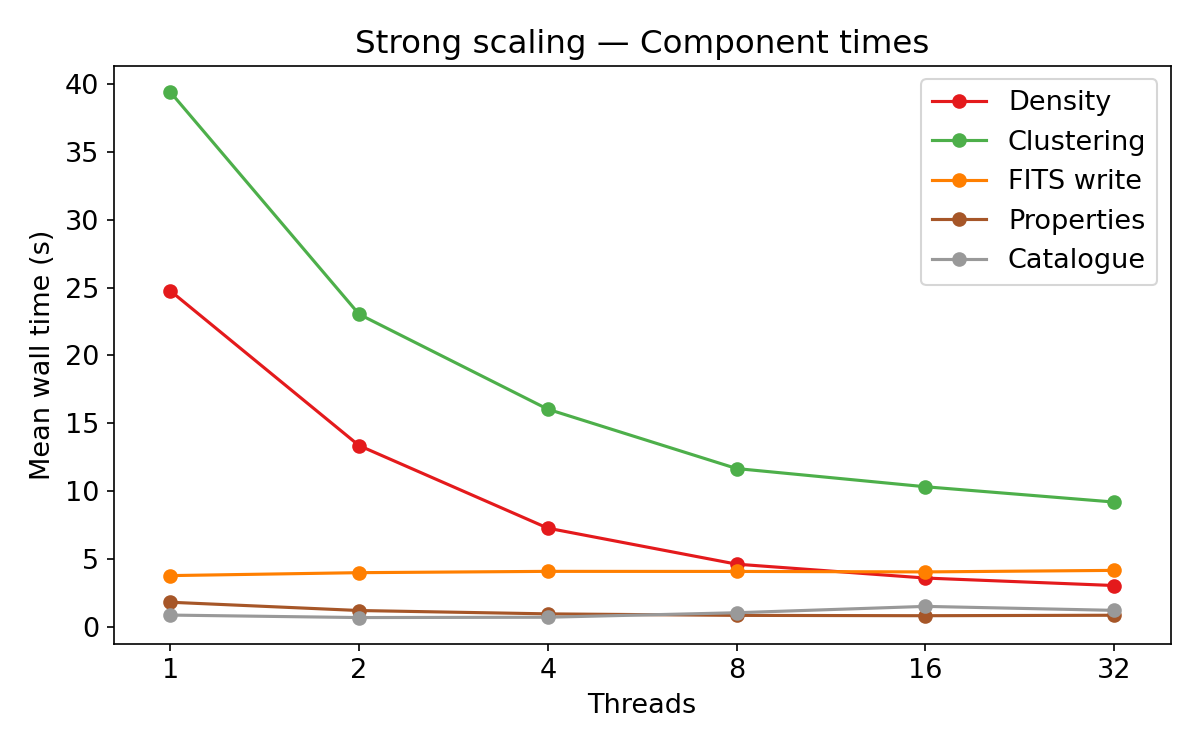}
    \caption{Mean execution time of the individual stages of the \ADP\ pipeline as a function of the number of threads in the strong-scaling experiments. The reported values correspond to the average over ten independent executions.}
    \label{fig:strong_time_breakdwon}
\end{figure}
\begin{figure}[t]
    \centering
    \includegraphics[width=0.5\textwidth]{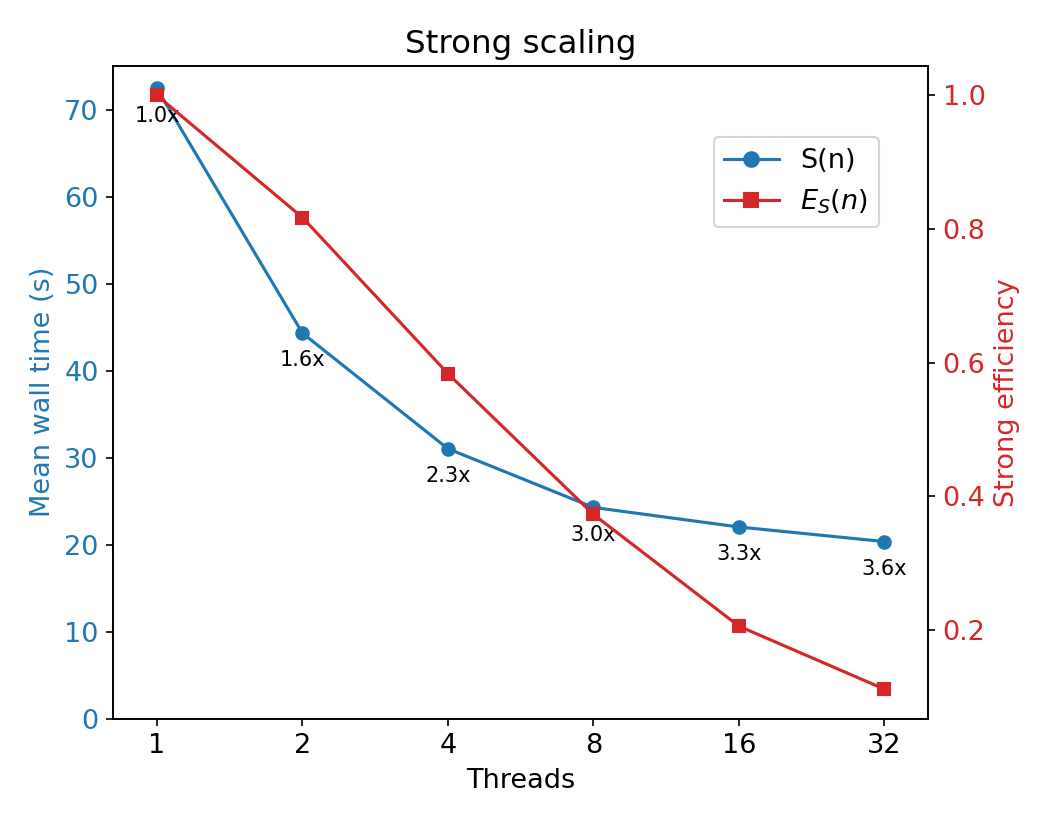}
    \caption{Wall-clock time and parallel efficiency as a function of the number of OpenMP threads for the strong-scaling experiments. The speedup values are reported above each data point. Both quantities are defined according to Eq.~\eqref{eq:parallel_efficiency_strong}.}
    \label{fig:strong_time_eff}
\end{figure}

Figure~\ref{fig:strong_time_breakdwon} reports the mean execution time of the individual computational stages, as a function of the number of OpenMP threads. The density estimation and clustering stages exhibit the strongest reduction in execution time, demonstrating that the computationally dominant components of \ADP\ effectively benefit from shared-memory parallelization. In contrast, the segmentation-map generation, source-property computation and catalog export exhibit only a weak dependence on the number of threads, as these stages are largely serial or dominated by I/O operations.

Figure~\ref{fig:strong_time_eff} shows the corresponding total wall-clock time together with the parallel efficiency $E_S(n)$. The speedup $S(n)$ is reported above each data point. The execution time decreases monotonically from approximately 70 s for a single thread to 20 s using 32 threads, corresponding to an overall speedup of about $3.6\times$. The speedup progressively departs from ideal linear scaling, with the parallel efficiency $E_S(n)$ decreasing toward larger thread counts. This behavior reflects the increasing relative contribution of serial components, together with synchronization overhead and the finite amount of independent work available within the processed image.

The latter contribution is particularly relevant for the patch-based parallelization adopted by \ADP. As individual detection patches are processed independently, the available parallelism depends on both their number and size. The scaling measured here is therefore also limited by the finite workload of the test image; larger or higher-resolution datasets, containing a greater amount of independent work per thread, may allow the available computational resources to be exploited more effectively.

\subsection{Weak scaling}

\begin{figure}[t]
    \centering
    \includegraphics[width=0.5\textwidth]{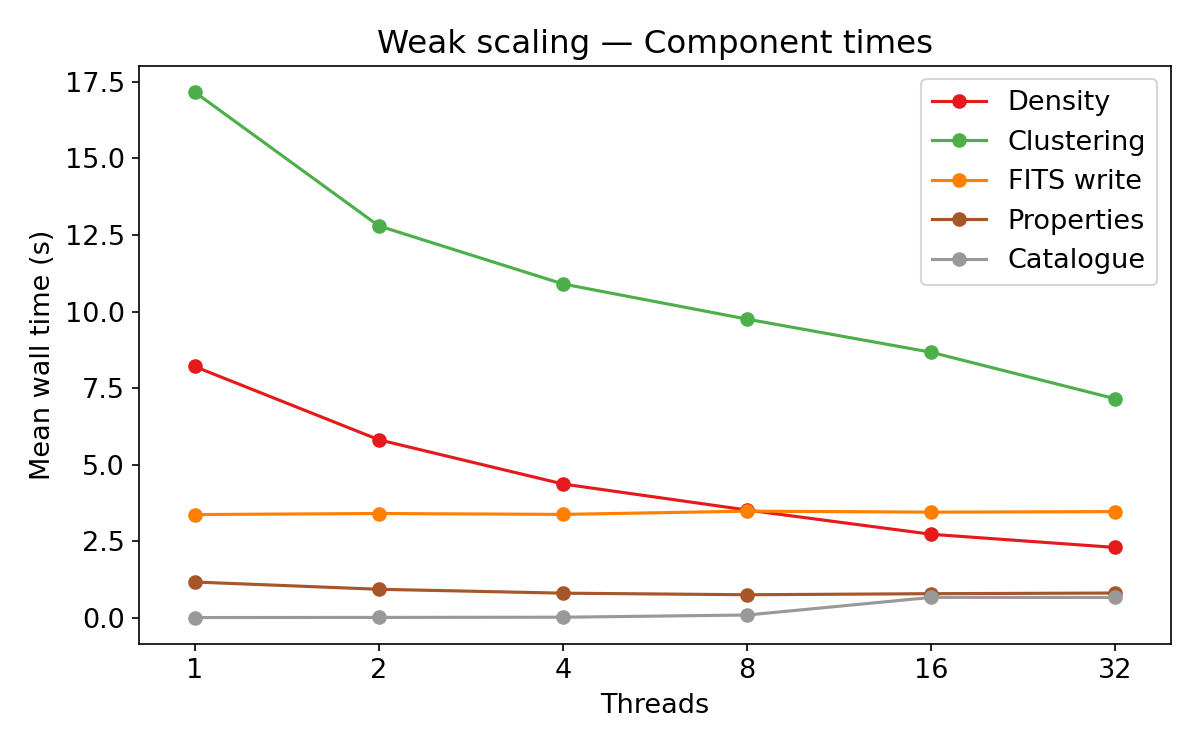}
    \caption{Mean execution time of the individual stages of the \ADP\ pipeline as a function of the number of threads in the weak-scaling experiments. The reported values correspond to the average over ten independent executions.}
    \label{fig:weak_time_breakdown}
\end{figure}

Similarly to the strong-scaling analysis, Figure~\ref{fig:weak_time_breakdown} reports the mean execution time of the individual computational stages, averaged over ten independent runs, as a function of the number of threads. Consistent with the strong-scaling results, density estimation and clustering account for the majority of the computational cost and show the largest variations across the explored configurations, while the remaining stages exhibit only modest variations. 
\begin{figure}[t]
    \centering
    \includegraphics[width=0.5\textwidth]{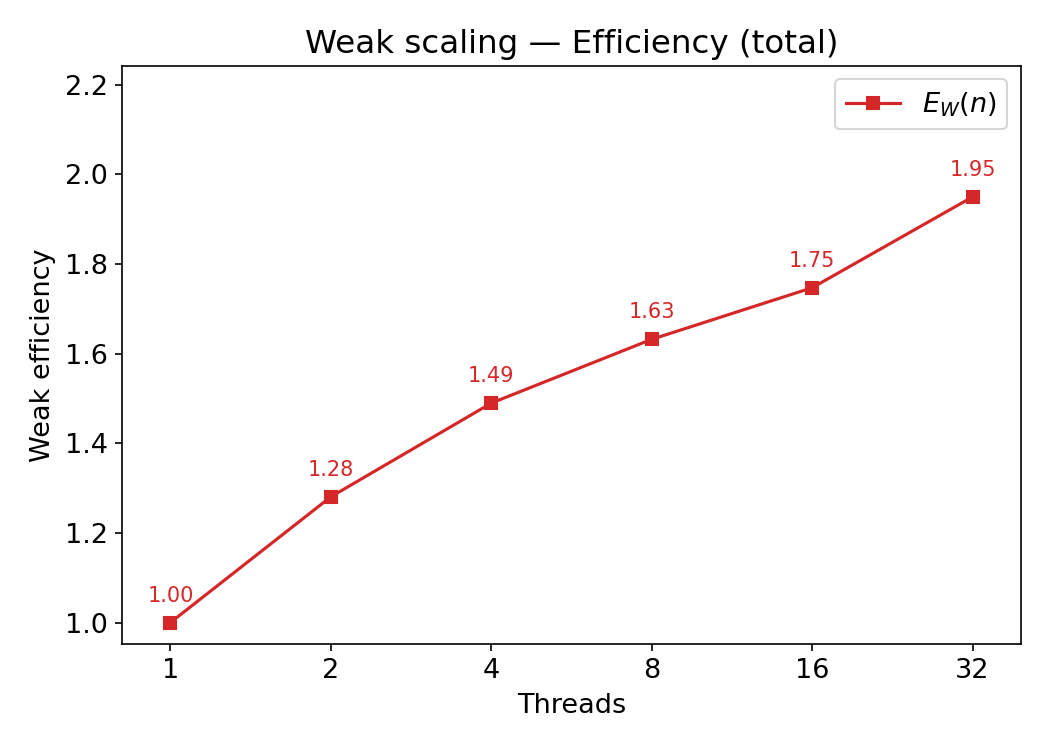}
    \caption{Wall-clock time and parallel efficiency as a function of the number
     of OpenMP threads for the weak-scaling experiments. The speedup values are not reported separately, since the weak-scaling efficiency corresponds directly to the speedup, as defined in Eq.~\eqref{eq:parallel_efficiency_weak}.}
    \label{fig:weak_time_eff}
\end{figure}
\begin{figure}[h!]
    \centering
    \includegraphics[width=0.5\textwidth]{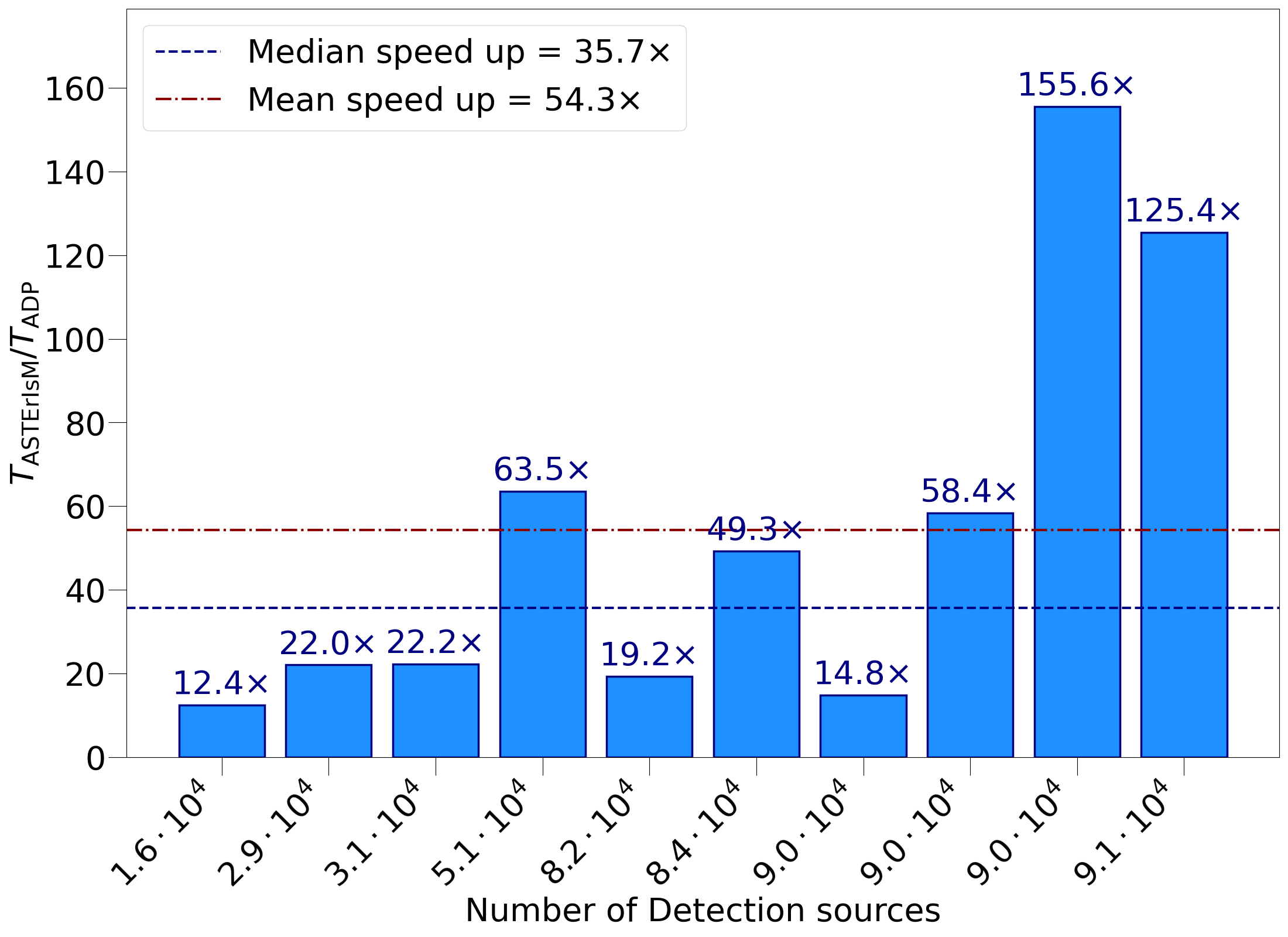}
    \caption{End-to-end speedup of \ADP\ relative to \aste\ for the selected EQ1 images, ordered by increasing number of detected sources. The speedup is defined as $T_{\mathrm{ASTErIsM}}/T_{\mathrm{ADP}}$ and the dashed horizontal lines mark the median and mean value across the sample. The number above each bar reports the corresponding speedup factor.}
    \label{fig:speed_up}
\end{figure}
Figure~\ref{fig:weak_time_eff} shows the weak-scaling efficiency $E_W(n)$. In contrast to ideal weak scaling, for which $E_W(n)=1$, the measured efficiency is systematically larger than unity and increases with the number of threads. Since $E_W(n)=T(1)/T(n)$ in the weak-scaling case (Eq.~\ref{eq:parallel_efficiency_weak}), this behavior reflects a decrease in execution time despite the approximately constant workload assigned to each compute unit. This apparent super-linear behavior suggests that the detection \texttt{ISOAREA} adopted as a workload proxy does not fully capture the computational complexity of individual patches. Moreover, the finite number and heterogeneous sizes of the detection patches make it difficult to construct strictly equivalent workloads per thread, particularly at larger thread counts. The present results should therefore be interpreted as an assessment of the behavior of the patch-based parallelization under increasing workloads rather than as a strict measurement of weak-scaling efficiency.

\subsubsection{Runtime comparison with \aste}
\label{run_time_results}
Figure~\ref{fig:speed_up} summarizes the execution times measured for \ADP\ and \aste\ on the selected sample of EQ1 images together with the corresponding speedup factor, using the configuration reported in Section~\ref{end-to-end-methodology}. Across the analyzed sample, \ADP\ consistently requires substantially less execution time than \aste. While the execution time of \aste\ typically ranges from several tens of minutes up to more than one hour per image, \ADP\ processes the same observations in approximately 15-200 s. The resulting speedup spans from about $12\times$ to $156\times$, with a median value of $35.7\times$ and a mean value of $54.3\times$. 

The speedup varies considerably across the analyzed images and does not exhibit a simple dependence on the number of detected sources. This behavior indicates that the computational cost of the two deblenders depends on additional properties of the analyzed fields beyond the source count. A detailed analysis of the runtime dependence on the characteristics of the detection segmentation maps is presented in~\ref{Appendix_runtime}. 

Despite the variation across individual images, the computational advantage of \ADP\ remains substantial across the entire sample. Combined with the comparable deblending performance demonstrated in the preceding analyses, these results show that \ADP\ can achieve similar scientific performance at a substantially lower computational cost than \aste.

\section{Conclusion}
\label{conclusion}
In this work, we presented a redesign of the Advanced Density Peak (\ADP) clustering algorithm tailored to astronomical source deblending, together with a modular validation framework that enables reproducible benchmarking on both simulated and real astronomical observations. 

The redesign combines a hierarchical density-based clustering strategy with the independent processing of detection regions, providing a computational framework that exposes both fine-grained algorithmic parallelism and coarse-grained patch-level parallelism, while retaining a limited number of interpretable parameters. The validation framework combines controlled pairwise experiments, synthetic multi-source images, automatically generated ground-truth (GT) segmentation maps, and real observations from the \textit{Euclid} Q1 (EQ1) public data release, providing a comprehensive benchmark for assessing both the scientific and computational performance of astronomical deblending algorithms.

Across the different validation stages, \ADP\ achieves deblending performance comparable to \aste, the density-based deblending algorithm currently adopted within the \textit{Euclid} data processing pipeline~\citep{Euclid:2025wfi}. Controlled pairwise simulations show similar deblending accuracy across the explored source separations and flux ratios, with minor difference in pairwise precision, pairwise recall, and the recovered source photometry. For the majority of configurations, the recovered fluxes agree to within approximately 1\%, with residuals reaching only $5-7\%$ in the most challenging blending scenarios. No systematic photometric offset is observed, with either deblender recovering the larger flux depending on the specific blending configuration (Figures~\ref{fig:pp_comparison}--\ref{fig:pair_phot_comparison}).

In the more complex multi-source simulations, \ADP\ recovers 5515 GT sources compared with 5101 for \aste, corresponding to approximately 8\% more matched sources, while the positions of the 5062 sources recovered by both methods agree at the subpixel level (Figures~\ref{fig:matching_distance_distribution} and~\ref{fig:matching_offsets}). 

The comparison on EQ1 public data release further demonstrates a high degree of consistency in the recovered morphology, namely \texttt{ISOAREA}, \texttt{ELLIPTICITY}, \texttt{POSITION\_ANGLE}, and photometric properties . While the agreement in \texttt{ISOAREA} and photometry is strongest for extended sources and gradually degrades toward the smallest and faintest detections, the scatter in \texttt{ELLIPTICITY} and \texttt{POSITION\_ANGLE} instead increases for larger and more elongated objects, reflecting their stronger sensitivity to differences in the reconstructed segmentation geometry (Figures~\ref{fig:q1_isoarea}--\ref{fig:q1_photometry}).

The computational analysis demonstrates that the scientific performance is achieved at substantially lower computational cost. The strong-scaling experiments show that the computationally dominant density-estimation and clustering stages effectively exploit the available shared-memory parallelism, although the overall efficiency progressively decreases with increasing thread count owing to the contribution of serial components and the finite parallelism available across independent detection patches. The weak-scaling analysis further demonstrates that the patch-based implementation maintains a favorable computational behavior under increasing workloads, while highlighting the challenges associated with balancing heterogeneous astronomical detection regions (Figures~\ref{fig:strong_time_breakdwon}--\ref{fig:weak_time_eff}). 

More importantly, the end-to-end comparison on EQ1 observations shows that \ADP\ is consistently faster than \aste, with speedups ranging from approximately $12\times$ to $156\times$ and a median improvement of $35.7\times$ across the analyzed sample (Figure~\ref{fig:speed_up}).

Taken together, these results establish \ADP\ as a competitive alternative for astronomical source deblending, combining segmentation and photometric performance comparable to \aste\ with a substantially reduced computational cost. Future work will extend the validation to a broader range of source morphologies and observational conditions, while further investigating the dependence of the computational workload on the properties of individual detection regions. 

Additionally, the redesign presented in this work provides a computational framework that naturally exposes parallelism at multiple levels, from the shared-memory parallelization of the algorithm itself to the independent processing of detection patches. This computational hierarchy forms the basis of the distributed-memory implementation, \DADP, enabling the extension of the current patch-based approach beyond a single compute node and providing a scalable framework for the efficient processing of increasingly large astronomical imaging datasets.

\section*{Acknowledgements}
This paper is supported by the Agenzia Spaziale Italiana (ASI) under - EUCLID-FASE E Attività scientifica per la missione - Accordo Attuativo n. 2024-10-HH.0. The code development, computational runs, and post-processing has been done on both LeonardoBooster platform, made available by the CINECA Italian national HPC facility through the EuroHPC joint undertaking (project EUHPC\_D31\_027), and PLEIADI, a computing infrastructure installed and managed by INAF-USCVIII. We would like to thank Andrea Zacchei for valuable discussions.

\bibliographystyle{elsarticle-harv} 
\bibliography{biblio}

\appendix
\newpage
\section{Runtime dependence on image properties}
\label{Appendix_runtime}

The end-to-end runtime comparison presented in Section~\ref{run_time_results} showed that the execution time of the two deblenders cannot be explained solely by the number of detected sources. This Appendix further investigates the computational behavior of \ADP\ and \aste\ by relating their execution times to several properties of the corresponding detection segmentation maps.

To provide a quantitative comparison of the observed trends, the execution times were fitted using a log-linear model of the form $\log_{10}(T)=ax+b$, where $T$ denotes the runtime and $x$ the image property under consideration. The quality of each fit was quantified through the coefficient of determination, $R^2$, defined as
\[
R^2 = 1 - \frac{\sum_i (y_i-\hat{y}_i)^2}
              {\sum_i (y_i-\bar{y})^2},
\]
where $y_i=\log_{10}(T_i)$, $\hat{y}_i$ is the corresponding fitted value, and $\bar{y}$ is the mean of the observed $\log_{10}(T)$ values. Values of $R^2$ approaching unity indicate that the fitted dependence accounts for most of the observed variation in $\log_{10}(T)$, whereas values approaching zero indicate that little of the observed variation is described by the model. Given the limited sample of ten EQ1 images, these values are intended only as descriptive indicators of the observed trends rather than as statistically rigorous estimates.

\begin{figure}[t]
    \centering
    \includegraphics[width=0.5\textwidth]{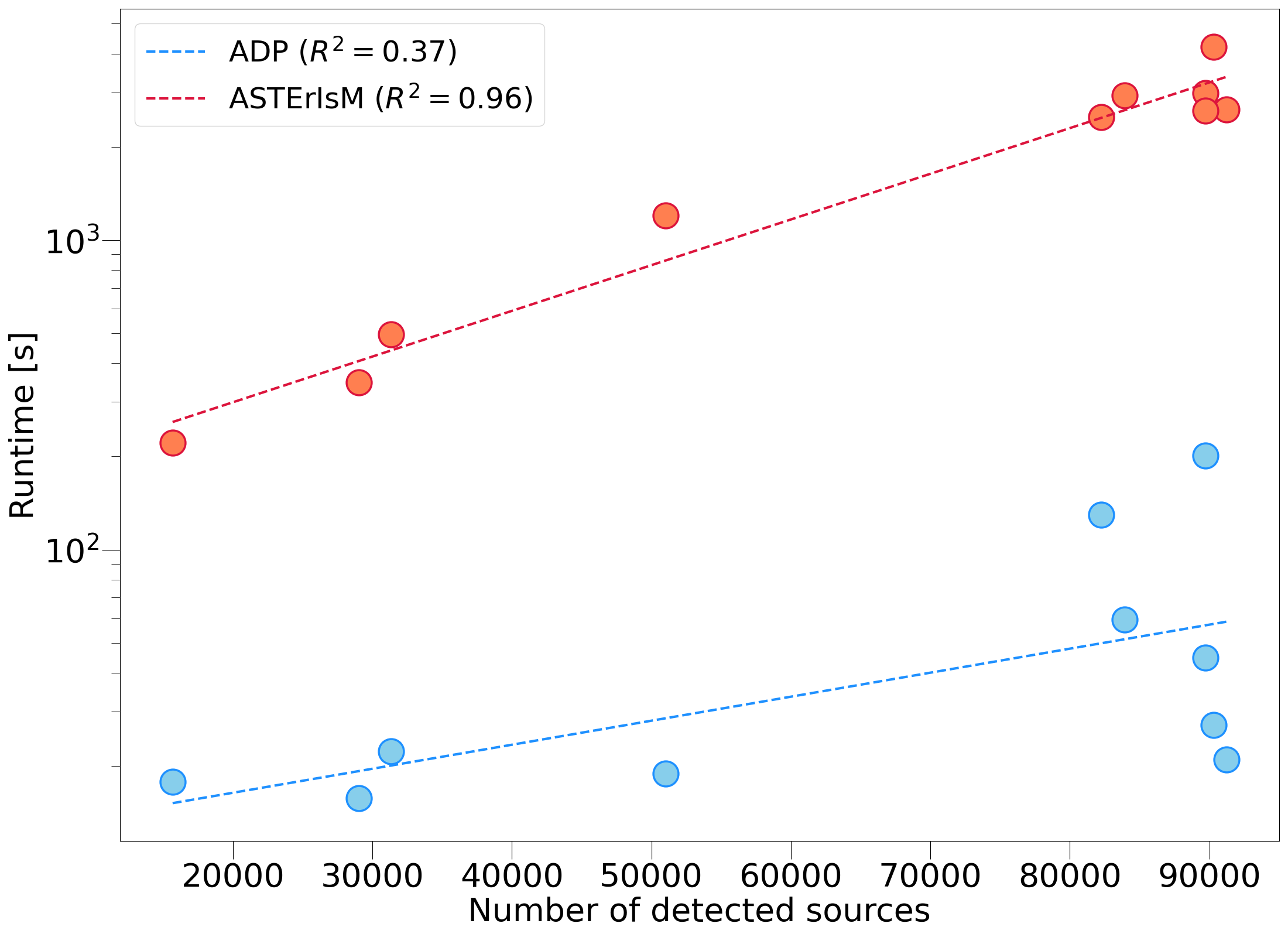}
    \caption{Execution time of \ADP\ and \aste\ as a function of the number of detected sources for the ten EQ1 images analyzed in this work. The dashed lines show the best-fitting log-linear models used to describe the observed runtime trends. The corresponding coefficients of determination $R^2$ evaluated are reported in the legend.}
    \label{fig:runtime_nsources}
\end{figure}

Figure~\ref{fig:runtime_nsources} compares the execution time of the two deblenders as a function of the number of detected sources. The runtime of \aste\ exhibits a strong correlation with the source population, whereas the corresponding dependence is substantially weaker for \ADP, indicating that the number of detected sources alone provides only a limited predictor of its computational cost.
\begin{figure}[h!]
    \centering
    \includegraphics[width=0.5\textwidth]{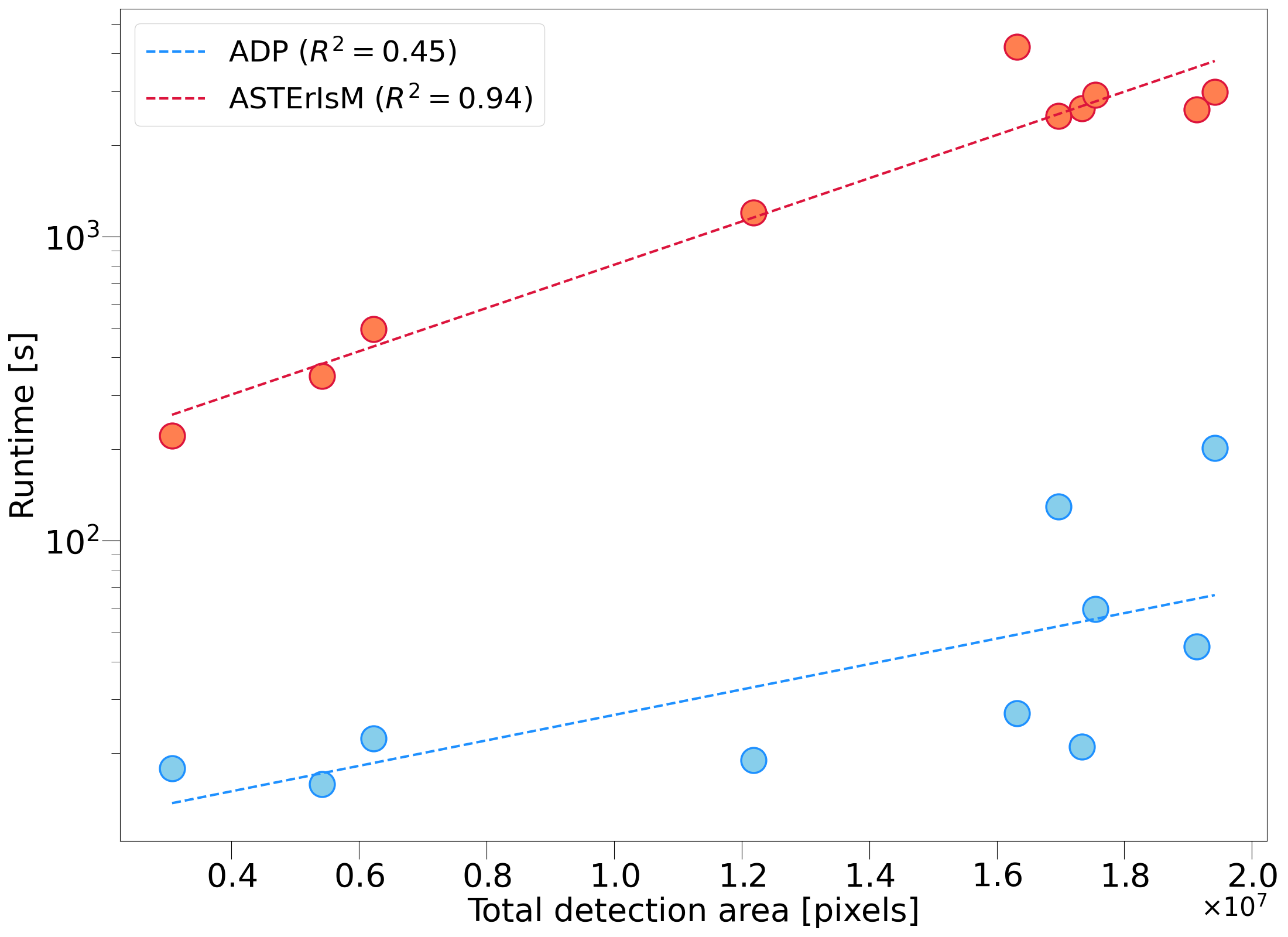}
    \caption{Execution time of \ADP\ and \aste\ as a function of the total detection area for the ten EQ1 images analyzed in this work. The dashed lines show the best-fitting log-linear models used to describe the observed runtime trends. The corresponding coefficients of determination $R^2$ evaluated are reported in the legend.}
    \label{fig:runtime_total_area}
\end{figure}
\begin{figure}[h!]
    \centering
    \includegraphics[width=0.5\textwidth]{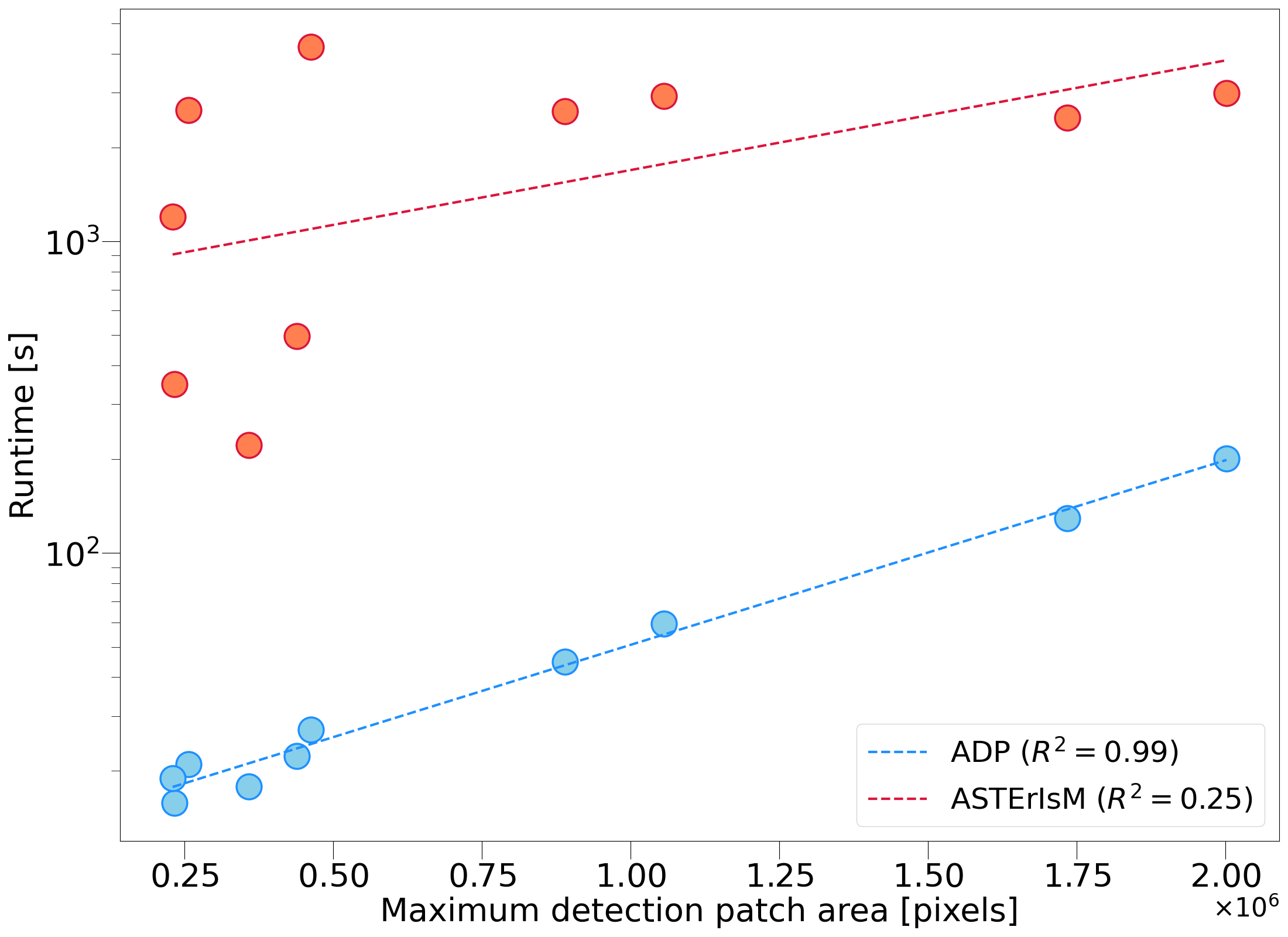}
    \caption{Execution time of \ADP\ and \aste\ as a function of the maximum detection-patch area for the ten EQ1 images analyzed in this work. The dashed lines show the best-fitting log-linear models used to describe the observed runtime trends. The corresponding coefficients of determination $R^2$ evaluated are reported in the legend.}
    \label{fig:runtime_max_area}
\end{figure}

Figure~\ref{fig:runtime_total_area} presents the corresponding comparison as a function of the total detection area. Similar to the previous case, the execution time of \aste\ remains strongly correlated with this global property of the analyzed field, whereas only a weak dependence is observed for \ADP. 

Finally, Figure~\ref{fig:runtime_max_area} relates the execution time to the maximum detection-patch area. In contrast to the previous comparisons, the runtime of \ADP\ exhibits an almost perfect correlation with the size of the largest detection patch, whereas no comparable dependence is observed for \aste.

\begin{figure*}[t]
    \centering
    \includegraphics[width=0.90\textwidth]{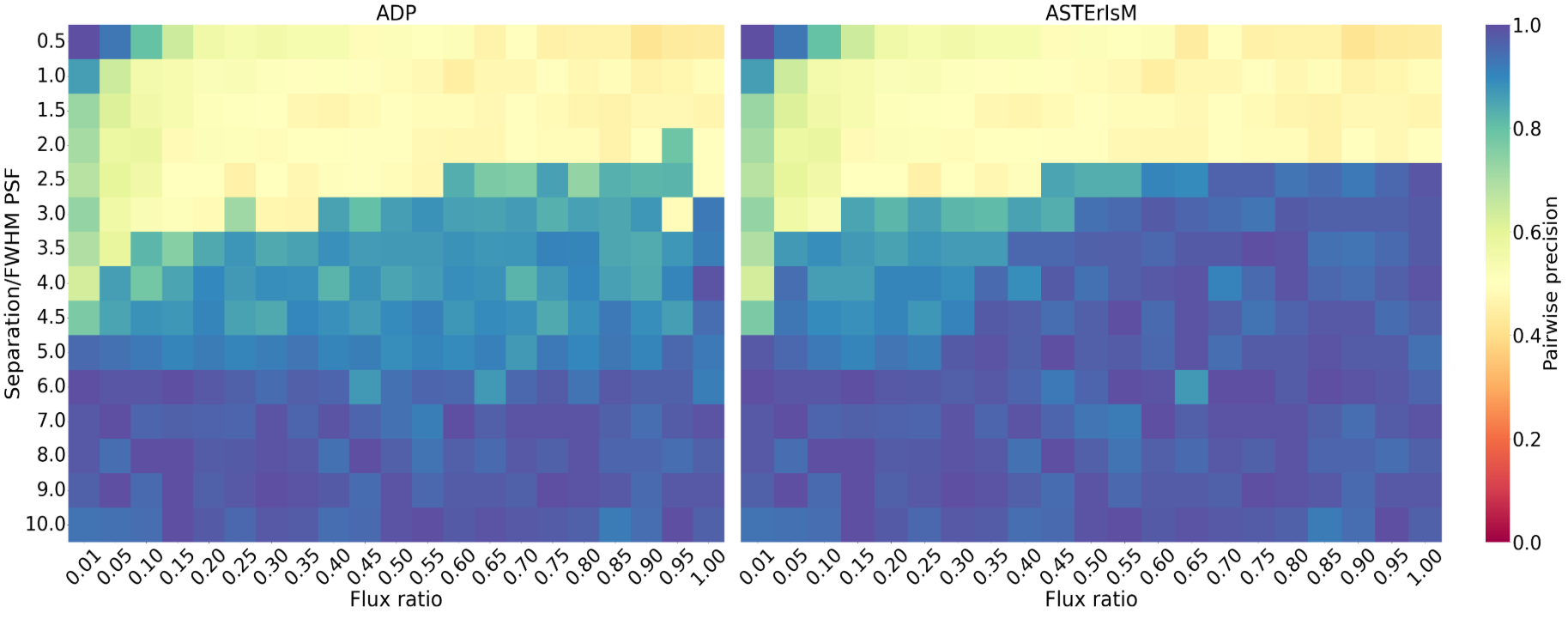}
    \includegraphics[width=0.90\textwidth]{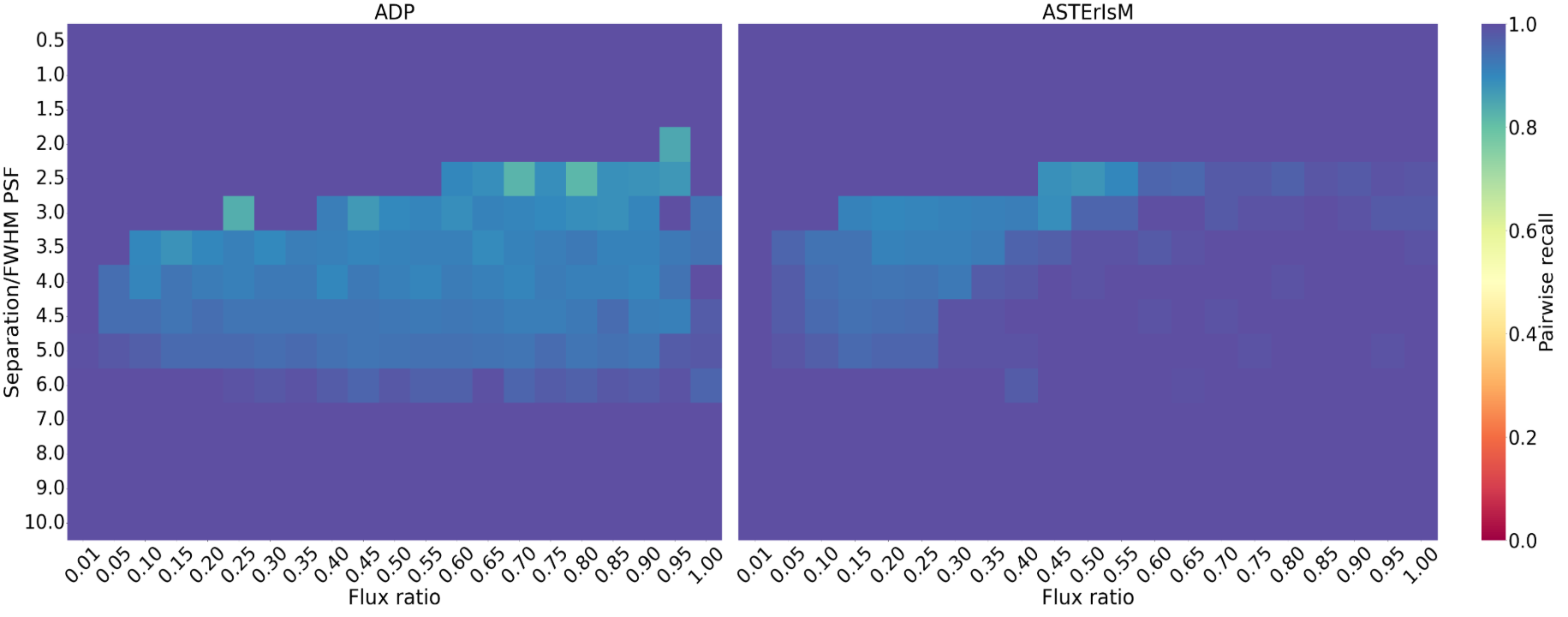}
    \caption{
    PP (top) and PR (bottom) as a function of source separation and flux ratio for the star-star pairwise simulations. In both cases, the left and right panels show the results obtained with \ADP\ and \aste, respectively.
    }
    \label{fig:pp_pr_comparison_star_star}
\end{figure*}

Table~\ref{tab:runtime_r2} summarizes the coefficients of determination obtained for the three image properties considered in this Appendix. The comparison clearly illustrates that the computational cost of \aste\ is best described by global properties of the analyzed field, whereas the runtime of \ADP\ is almost entirely determined by the maximum detection-patch area. This behavior is fully consistent with the patch-based execution strategy adopted by \ADP, confirming that its computational complexity is primarily governed by the largest independently processed detection region rather than by the global properties of the analyzed image. These results demonstrate that the computational cost of the two deblenders is governed by fundamentally different image properties, reflecting the distinct algorithmic strategies adopted by the two methods.

\section{Controlled pairwise simulations: star-star and star-galaxy}
\label{Controlled_pairwise_additional}
\begin{table}[t]
\centering
\caption{Coefficient of determination $R^2$, evaluated in $\log_{10}(\mathrm{T_i})$ space, for the log-linear fits relating the execution time of the two deblenders to different properties of the detection segmentation maps.}
\label{tab:runtime_r2}
\begin{tabular}{lcc}
\hline
Image property & $R^2_{\ADP}$ & $R^2_{\aste}$ \\
\hline
Number of detected sources & 0.37 & 0.96 \\
Total detection area & 0.45 & 0.94 \\
Maximum detection-patch area & 0.99 & 0.25 \\
\hline
\end{tabular}
\end{table}
\begin{figure*}[t]
    \centering
    \includegraphics[width=0.90\textwidth]{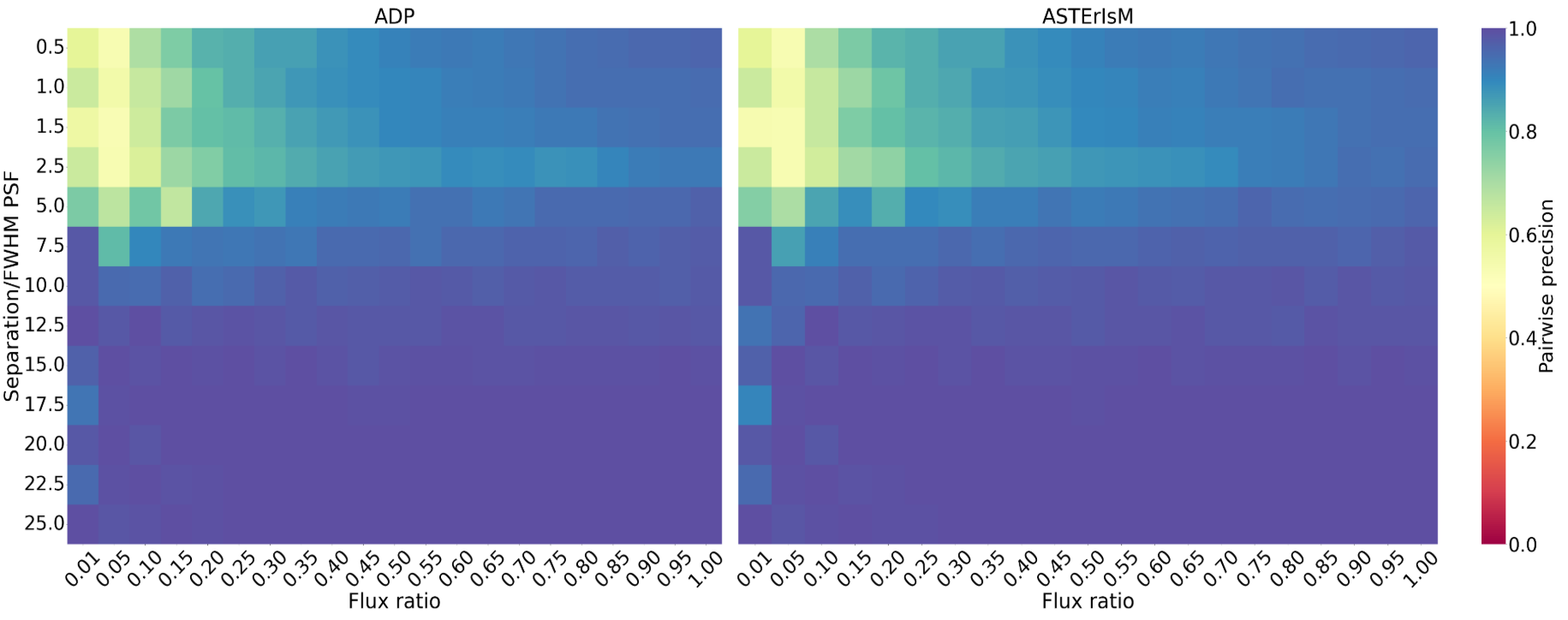}
    \includegraphics[width=0.90\textwidth]{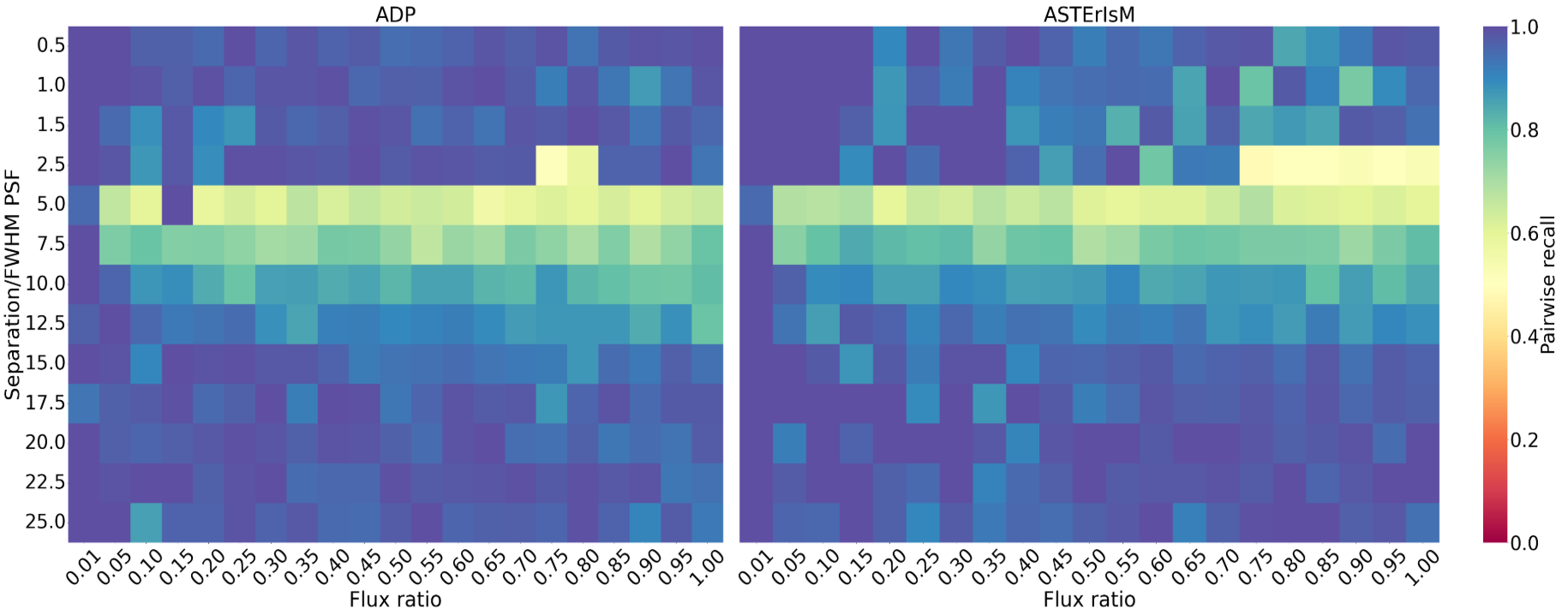}
    \caption{
     PP (top) and PR (bottom) as a function of source separation and flux ratio for the star-galaxy pairwise simulations. In both cases, the left and right panels show the results obtained with \ADP\ and \aste, respectively.
    }
    \label{fig:pp_pr_comparison_star_gal}
\end{figure*}
\begin{figure*}[h!]
    \centering
        \includegraphics[width=1.0\textwidth]{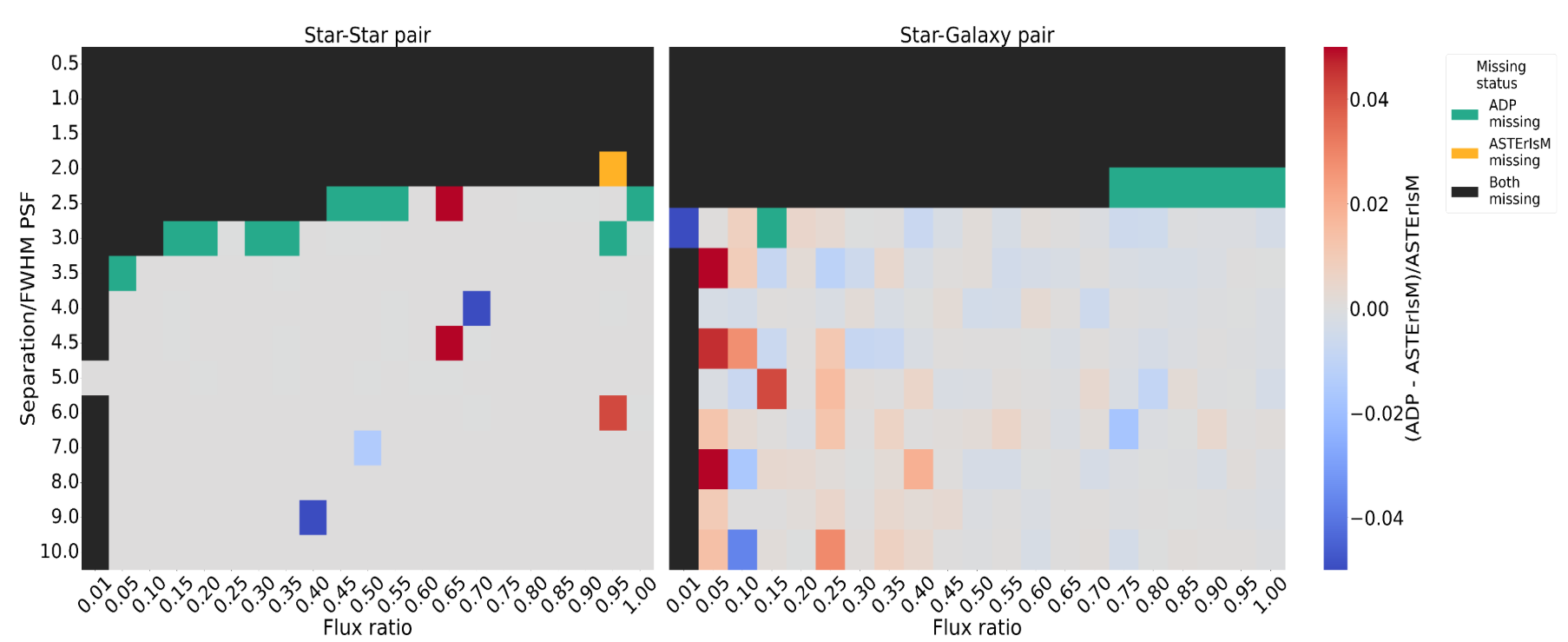}
    \caption{
    Residuals of \texttt{FLUX\_{ISOAREA}} measured by \ADP\ and \aste\ for the matched sources, as a function of source separation and flux ratio. The left and right panels show the star-star and star-galaxy simulations, respectively. Dedicated colors highlight configurations in which one source is missed by \ADP, one source is missed by \aste, or one source is missed by both methods.}
    \label{fig:pair_phot_comparison_star_star_gal_gal}
\end{figure*}
This Appendix complement the analysis presented in Section~\ref{controlled_pairwise_gal_gal} by reporting the results for the star--star and star--galaxy simulations. The same validation procedure described in Sections~\ref{Synthetic_images} and~\ref{Pairwise_metrics} is adopted throughout, including the PP and PR~\eqref{eq:pairwise_metrics} evaluation and the photometric comparison based on the \texttt{FLUX\_ISOAREA} estimator.

Figure~\ref{fig:pp_pr_comparison_star_star} presents the PP and PR obtained for the star--star simulations, while Figures~\ref{fig:pp_pr_comparison_star_gal} shows the corresponding results for the star--galaxy case. In both configurations, \ADP\ and \aste\ exhibit behavior fully consistent with that observed for the galaxy--galaxy simulations, with only minor differences appearing in the most strongly blended configurations. 

The corresponding photometric comparisons are shown in Figure~\ref{fig:pair_phot_comparison_star_star_gal_gal}, where the star--star and star--galaxy results are presented side by side. As in the galaxy--galaxy case, the recovered photometry exhibits excellent agreement between the two deblenders, confirming that the small differences in the reconstructed segmentation translate into only minor variations in the recovered source fluxes. No systematic photometric offset is observed, with either deblender recovering the larger flux depending on the specific blending configuration.

\section{From the ideal to the detection-limited ground truth}
\begin{figure}[h!]
\centering
\includegraphics[width=0.5\textwidth]{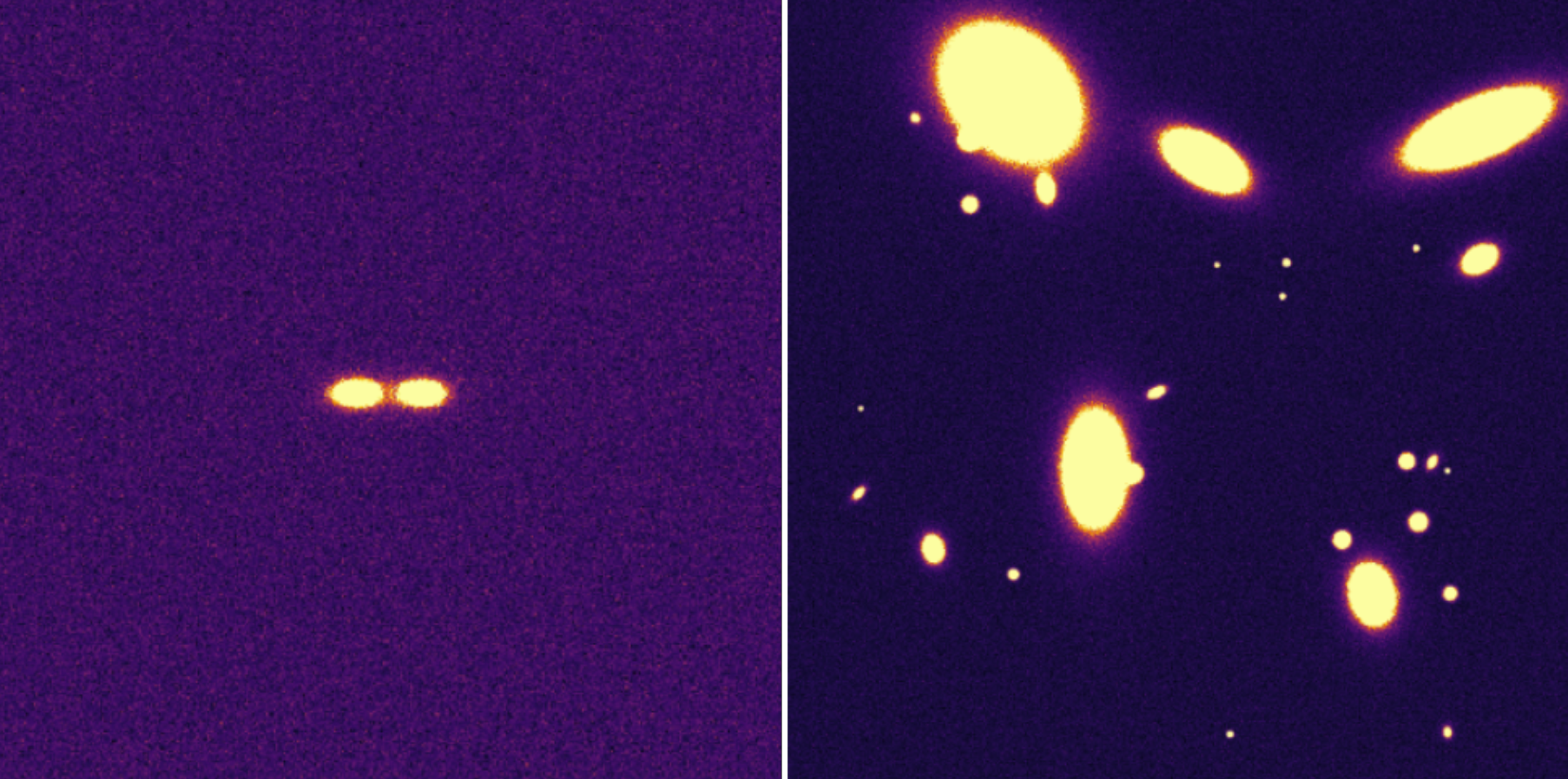}

\vspace{0.1cm}

\includegraphics[width=0.5\textwidth]{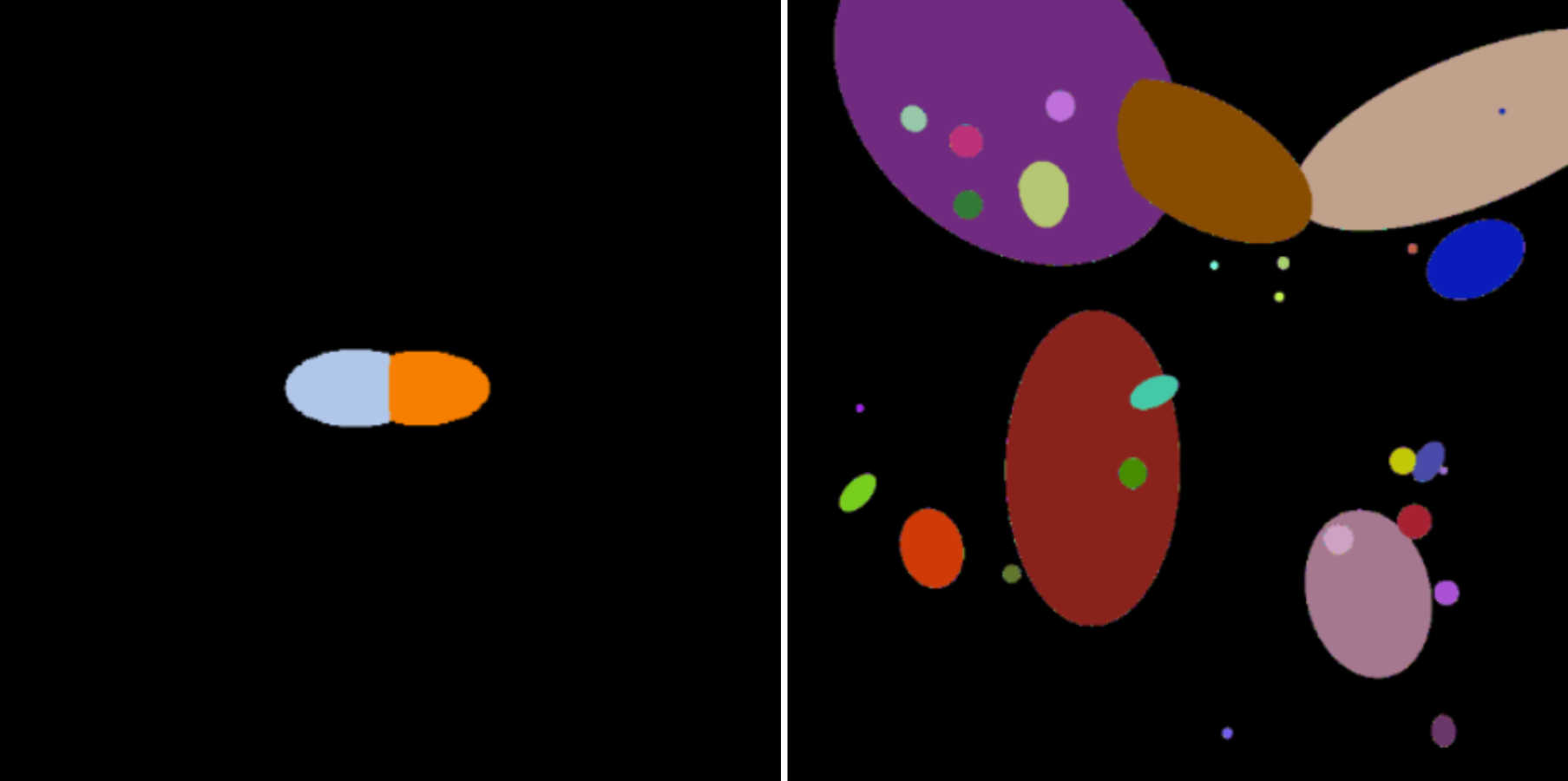}

\vspace{0.1cm}

\includegraphics[width=0.5\textwidth]{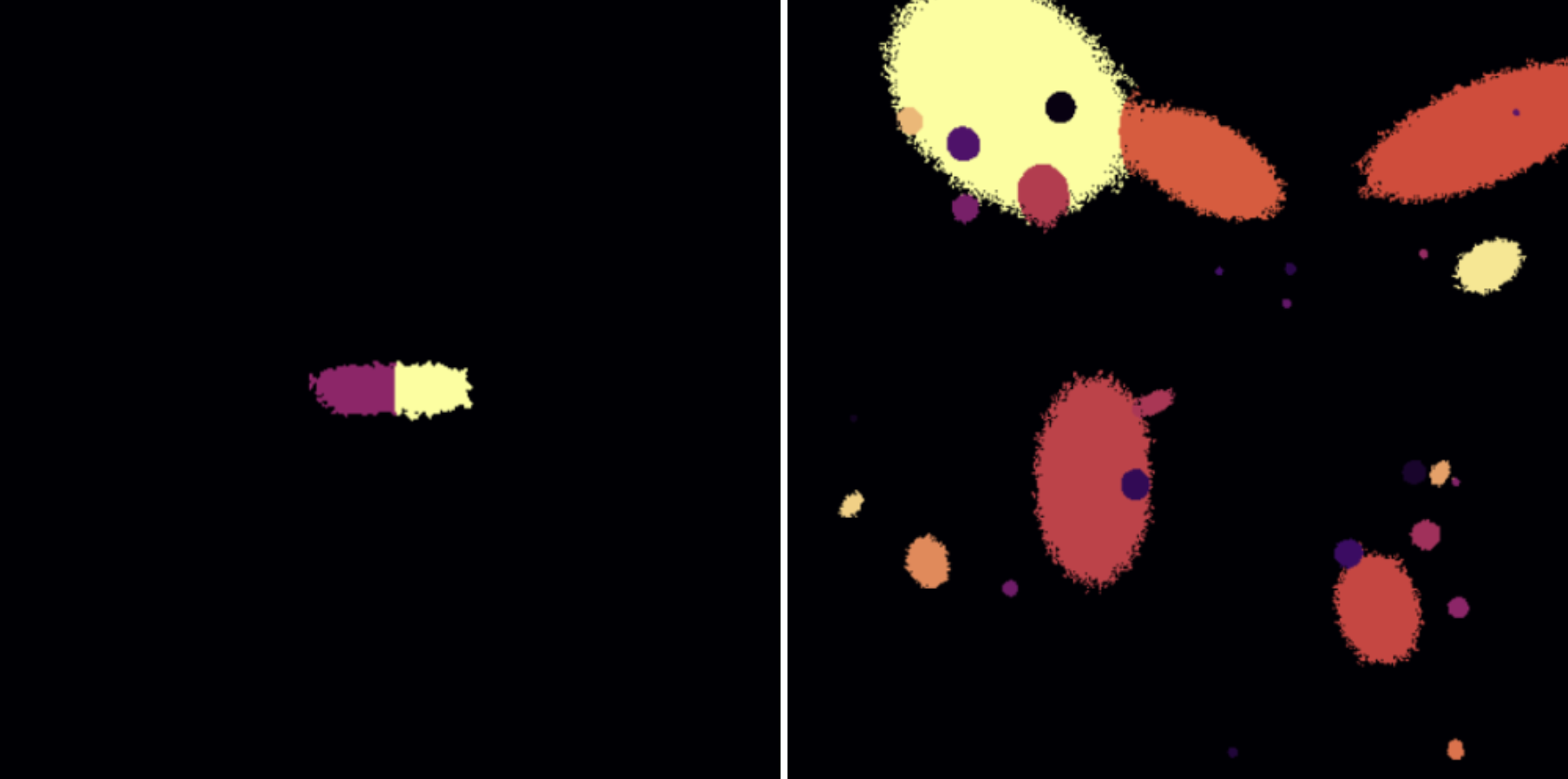}

\caption{Construction of the reference GT segmentation for a galaxy-galaxy pair (left) and a synthetic multi-source image (right). From top to bottom: simulated image, ideal GT, and reference GT after masking with the \se\ detection segmentation map.}
\label{fig:GT_creation_example}
\end{figure}

The GT segmentation adopted throughout this work is obtained through the procedure described in Section~\ref{final_GT_creation}. For completeness, Figure~\ref{fig:GT_creation_example} illustrates the different stages of its construction for both a controlled galaxy--galaxy pair and a synthetic multi-source image. Starting from the simulated image, an ideal GT segmentation is first generated by assigning each pixel to the source providing the largest significant contribution to its flux. The final reference GT is then obtained by masking this ideal segmentation with the \se\ detection segmentation map, thereby reproducing the detection limitations of a realistic astronomical pipeline while preserving the unambiguous pixel assignment of the ideal GT. The resulting reference GT segmentation constitutes the common benchmark adopted throughout all the validation experiments presented in this work

\end{document}